\documentclass[twocolumn]{aastex62}

\usepackage{verbatim}

\newcommand{\kms}{km~s$^{-1}$} 
\newcommand{\water}{H$_2$O} 
\newcommand{\methanol}{CH$_3$OH}

\newcommand{\lsun}{$L_\odot$}
\newcommand{\msun}{$M_\odot$}

\newcommand{\ngc}{NGC\,6334}
\newcommand{\ngci}{{\ngc}I}

\newcommand{\mjb}{mJy~beam$^{-1}$}
\newcommand{\jb}{Jy~beam$^{-1}$}

\newcommand{\ccc}{cm$^{-3}$}
\newcommand{\cc}{cm$^{-3}$}
\newcommand{\tdust}{$T_{dust}$}

\newcommand{\fshift}{$f_{\rm shift}$}

\shorttitle{Extraordinary Outburst in \ngci-MM1: \water\/ masers}
\shortauthors{Brogan et al.}

\begin{document}

\title{The extraordinary outburst in the massive protostellar system \ngci-MM1:\\ Flaring of the water masers in a north-south bipolar outflow driven by MM1B}


\author{C. L. Brogan}
\affiliation{NRAO, 520 Edgemont Rd, Charlottesville, VA 22903, USA} 

\author{T. R. Hunter}
\affiliation{NRAO, 520 Edgemont Rd, Charlottesville, VA 22903, USA}

\author{C. J. Cyganowski}
\affiliation{SUPA, School of Physics and Astronomy, University of St. Andrews, North Haugh, St. Andrews KY16 9SS, UK}

\author{J. O. Chibueze}
\affiliation{Department of Physics and Astronomy, University of Nigeria, Carver Building, 1 University Road, Nsukka, 410001, Nigeria}
\affiliation{SKA South Africa, 3rd Floor, The Park, Park Road, Pinelands, Cape Town, 7405, South Africa}
\affiliation{Centre for Space Research, Physics Department, North-West University, Potchefstroom, 2520, South Africa}

\author{R. K. Friesen}
\affiliation{NRAO, 520 Edgemont Rd, Charlottesville, VA 22903, USA}

\author{T. Hirota}
\affiliation{National Astronomical Observatory of Japan, Osawa 2-21-1, Mitaka, Tokyo 181-8588, Japan}

\author{G. C. MacLeod}
\affiliation{Hartebeesthoek Radio Astronomy Observatory, PO Box 443, Krugersdorp 1740, South Africa}
\affiliation{The University of Western Ontario, 1151 Richmond Street, London, ON N6A 3K7, Canada}

\author{B. A. McGuire}
\altaffiliation{B.A.M. is a Hubble Fellow of the National Radio Astronomy Observatory}
\affiliation{NRAO, 520 Edgemont Rd, Charlottesville, VA 22903, USA}

\author{A. M. Sobolev} 
\affiliation{Astronomical Observatory, Institute for Natural Sciences and Mathematics, Ural Federal University, Ekaterinburg, 620000, Russian Federation}

\correspondingauthor{C. L. Brogan}
\email{cbrogan@nrao.edu}

\begin{abstract}  
We compare multi-epoch sub-arcsecond VLA imaging of the 22~GHz water masers toward the massive protocluster \ngci\/ observed before and after the recent outburst of MM1B in (sub)millimeter continuum.  Since the outburst, the water maser emission toward MM1 has substantially weakened.  Simultaneously, the strong water masers associated with the synchrotron continuum point source CM2 have flared by a mean factor of 6.5 (to 4.2~kJy) with highly-blueshifted features (up to 70~\kms\/ from LSR) becoming more prominent. The strongest flaring water masers reside 3000~au north of MM1B and form a remarkable bow shock pattern whose vertex coincides with CM2 and tail points back to MM1B.  Excited OH masers trace a secondary bow shock located $\sim$120~au downstream. ALMA images of CS (6-5) reveal a highly-collimated north-south structure encompassing the flaring masers to the north and the non-flaring masers to the south seen in projection toward the MM3-UCHII region. Proper motions of the southern water masers over 5.3~years indicate a bulk projected motion of 117~\kms\/ southward from MM1B with a dynamical time of 170~yr.  We conclude that CM2, the water masers, and many of the excited OH masers trace the interaction of the high velocity bipolar outflow from MM1B with ambient molecular gas.  The previously-excavated outflow cavity has apparently allowed the radiative energy of the current outburst to propagate freely until terminating at the northern bow shock where it strengthened the masers. Additionally, water masers have been detected toward MM7 for the first time, and a highly collimated CS (6-5) outflow has been detected toward MM4.

\end{abstract}


\section{Introduction}

Episodic accretion onto protostars has been identified as a common phenomenon in star formation. Observational evidence for episodic accretion in the form of long-term (month-long to decade-long) luminosity outbursts extends across a wide range of protostellar mass, from below a solar mass \citep[HOPS~383,][]{Safron15} up to $\sim20$~\msun\/ \citep[S255IR-NIRS3,][]{Caratti17}.  
Variability surveys at infrared wavelengths
\citep{Minniti10,Lucas08} and submillimeter wavelengths \citep{Yoo17} continue to identify new events, primarily
toward low mass protostars, from Class~0 to Class~II.  
Recent accretion outbursts have also been inferred as an explanation for
measurements of the CO snow line at anomalously large
radii in low mass protostellar disks \citep{Hsieh18,Frimann17,Jorgensen13}.
In addition, the prevalence of episodic accretion has been invoked to explain the observed luminosity spread of protostars in clusters \citep[see][and references therein]{Jensen18}, but whether this phenomenon has a predominant effect remains in question \citep{Fischer17}.  Nevertheless, the impact of such outbursts on the surrounding protocluster, particularly from large accretion events in high-mass protostars, is potentially a very important feedback mechanism in cluster formation \citep{Lomax18,Stamatellos12}.

Another eruptive phenomenon associated with sites of high mass star formation is maser flares, such as the three past events observed in the water masers in the vicinity of Orion~KL \citep{Abraham81,Omodaka99,Hirota14}. 
The repeating nature of the Orion maser outbursts \citep{Tolmachev11} as well as the periodic features seen toward several high mass protostellar objects (HMPOs) in the 6.7~GHz methanol maser line \citep[e.g.][]{Stecklum17,Goedhart04}, have suggested that variations in the underlying protostar could be responsible for maser flares and maser variability.  A connection between water masers and protostellar outflows has been long established \citep{Tofani95}. But in the past decade, solid evidence for water masers tracing bipolar outflows from intermediate to high-mass protostars has been obtained via proper motion studies: some examples include S255IR-SMA1 \citep{Burns16},
LkH$\alpha$~234 \citep{Torrelles14},
IRAS~20126+4104, \citep{Moscadelli11},
and Cepheus~A~HW3d \citep{Chibueze12}.
At the same time, millimeter imaging of thermal molecular gas in bipolar outflows shows evidence for successive ejection events from both low-mass protostars \citep{Chen16,Plunkett15} and high-mass protostars \citep[HH80-81]{Qiu09}.  Recently, these two phenomena have been linked by the observation of multiple bow shocks traced by water masers driven by the massive protostar AFGL~5142~MM1 \citep{Burns17}, motivating the importance of obtaining complementary datasets of thermal gas and maser gas surrounding massive protostars.

\ngci\/ is a relatively nearby example \citep[$d=1.30\pm0.09$~kpc,][]{Chibueze14,Reid14} of a
cluster of massive protostars \citep{Hunter06} that is so deeply-embedded that the two strong hot core millimeter sources \citep[MM1 and MM2,][]{McGuire17,Zernickel12} remain undetected in sub-arcsecond mid-infrared images \citep{DeBuizer02}.  In our initial Atacama Large Millimeter Array (ALMA) observations of \ngci\/ at epoch 2015.6 \citep{Brogan16}, we resolved MM1 into seven components with typical separations of $\approx$1000~au and brightness temperatures ranging from 100-260~K, consistent with the gas temperatures implied by the presence of the high $K$ lines of the CH$_3$CN $J$=(13-12) and (12-11) ladders and copious \water\/ maser emission.   Comparison with earlier SMA images from 2008 showed an unexpectedly large increase in millimeter continuum flux density from MM1 while the other sources were unchanged, including the MM3-UCHII region \citep[previously known as NGC6334F,][]{Rodriguez82}, the other hot core MM2, and the massive but line-poor dust core MM4.  Subsequent observations with the Atacama Large Millimeter/Submillimeter Array (ALMA) confirmed that the increase had been sustained for over a year and was centered on the hypercompact (HC) HII region MM1B \citep{Hunter17}, while long-term single-dish maser monitoring revealed a remarkable flare in multiple species and transitions beginning in January 2015 \citep{Macleod18}. Recent followup observations of the 6.7~GHz Class II methanol transition with the Karl G. Jansky Very Large Array (VLA) confirmed that the flaring masers arise predominantly from the MM1 region \citep{Hunter18}.  Together, these pieces of evidence point to an accretion outburst similar to those predicted by hydrodynamic simulations of massive star formation \citep{Meyer17} as well as low mass star formation \citep{Vorobyov05}.  The scale and ongoing nature of the outburst from MM1B strongly suggest that it could represent the massive protostellar equivalent of an FU~Orionis event \citep{Larson03}.

In this paper, we present follow-up observations with the VLA in 1.3~cm \water\/ maser emission.  We find remarkable flaring in the \water\/ maser emission coinciding with the synchrotron source CM2, $2.2\arcsec$ (3000 au) north of MM1B, and proper motions in the water masers south of MM1B in the southward direction.  The striking agreement between the locations of the masers and the morphology of a dynamically young north-south bipolar outflow identified in ALMA images of thermal dense gas traced by CS (6-5) yields a consistent picture of an energetic outflow driven by the massive protostar MM1B as it penetrates the surrounding dense material.  We conclude that the previously-excavated outflow cavity has allowed the radiative energy of the current outburst to propagate to the ends of the flow where it has strengthened the maser emission, particularly at the northern bow shock.

\begin{deluxetable*}{lccc}
\tablewidth{0pc}
\tablecaption{VLA observing parameters \label{obs}}  
\tablehead{
\colhead{Parameter} & 
\colhead{Epoch 1} & \colhead{Epoch 2} & \colhead{Epoch 3}}
\startdata
Project Code & 493\_1 & 16B-402 & 17B-258 \\
Observing Date & 2011 Sep 09 & 2017 Jan 08  & 2017 Oct 08  \\
Mean Epoch   & 2011.7 & 2017.0  & 2017.8 \\
Antenna Configuration & A & A  &  B   \\
Time on Source (minutes) & 43  & 41  & 49     \\
FWHM Primary Beam & $2\arcmin$  &  $2\arcmin$ &  $2\arcmin$   \\
Polarization products & dual circular & dual circular & dual circular \\
Gain calibrator & J1717-3342 & J1717-3342 & J1717-3342 \\
Bandpass calibrator & J1924-2914 & J1924-2914 & J1924-2914  \\
Flux calibrator & 3C286 & 3C286 & 3C286  \\              
Bandwidth (\kms\/)  & 222 & 215  & 215  \\ 
Native channel width (\kms\/)\tablenotemark{a} & 0.21 & 0.105 & 0.105 \\
Spectral resolution (\kms\/) & 0.42 & 0.21 & 0.21 \\
Cube channel width (\kms\/) & 0.25  &  0.25 &  0.25   \\
Synthesized beam ($\arcsec\times\arcsec$ (P.A.$\arcdeg$)) & 
$0\farcs41\times0\farcs22 (-24)$  & $0\farcs30\times0\farcs23$ (-18) & $0\farcs79\times0\farcs22 (-6)$  \\
RMS noise: min, median, max (\mjb\/)\tablenotemark{b}  & 16, 24, 680  & 8, 12, 1300  & 4, 6, 990 \\
Relative position uncertainty [RA ($\arcsec$), Dec ($\arcsec$)]\tablenotemark{c} & 0.019, 0.034 & 0.020, 0.025 & 0.019, 0.066
\enddata

 \tablenotetext{a}{The native spectral resolution is two times larger than the native channel width due to the application of Hanning smoothing before calibration.}
 \tablenotetext{b}{The rms noise varies significantly with channel number due to dynamic range limitations; the first number (the minimum) is the most representative value for the noise values measured in empty channels.}
 \tablenotetext{c}{The relative position uncertainty in Right Ascension and Declination for a $6\sigma$ maser detection threshold.}
\end{deluxetable*}

\section{Observations}

\subsection{VLA Observations}
\label{vlaobs}

The VLA 22.235~GHz water maser spectral line data presented in this paper arise from three epochs: 2011.7, 2017.0, and 2017.8, as described in Table~\ref{obs}. The 2011.7 epoch data were also presented in \citet{Brogan16}. The K-band 1.3~cm continuum from the two newer epochs will be the subject of a forthcoming paper. The C-band 5~cm continuum data from epoch 2016.9 used in this paper were originally described in \citet{Hunter18}.  The two new epochs of 1.3~cm data were calibrated using the VLA pipeline\footnote{See 
\url{https://science.nrao.edu/facilities/vla/data-processing/pipeline/scripted-pipeline} for more information.} in the Common Astronomy Software Applications (CASA) package. The VLA
pipeline applies Hanning smoothing to the data which reduces ringing from strong spectral features and coarsens
the spectral resolution to twice the observed channel width. A strong maser feature was used to iteratively self-calibrate the spectral line data for each dataset. 
Stokes I cubes of the maser emission were made with a channel spacing (and effective spectral resolution) of 0.25~\kms. In order to better match the 2017 \water\/ data, we re-imaged the \water\/ maser data from September 2011 in order to create cubes sampled with narrower channels (0.25~\kms\/ spacing, with 0.40~\kms\/ effective spectral resolution) than our previous analysis in \citet{Brogan16}.  The resulting noise in a typical 2011.7 epoch signal-free channel is $\sim 20$~\mjb. The angular resolution achieved for each epoch is given in Table~\ref{obs}.  The full width half maximum (FWHM) of the VLA primary beam at 22.235~GHz is $2'$. Primary beam correction was applied to correct the extracted maser flux densities.

Due to the high peak signal strength of the water maser emission and the fact that it occupies much of the bandwidth of the high-resolution spectral window employed, a few channels of the maser data (from $-71$ to $-59$~\kms) are affected by spectral artifacts due to phase serialization in the correlator and usage of the default values of the local oscillator modulation frequencies \citep[\fshift,][]{Sault13}.  Fortunately, only the two strongest maser positions (near CM2) show contamination, which takes the form of $-36$~dB ghosts of the peak signals in the -17 to -5~\kms\/ range of channels, i.e. shifted by one quarter of the spectral window bandwidth (54~\kms).  We have removed these false components from our analysis.



\begin{figure*}[ht!]  
\includegraphics[width=1.0\linewidth]{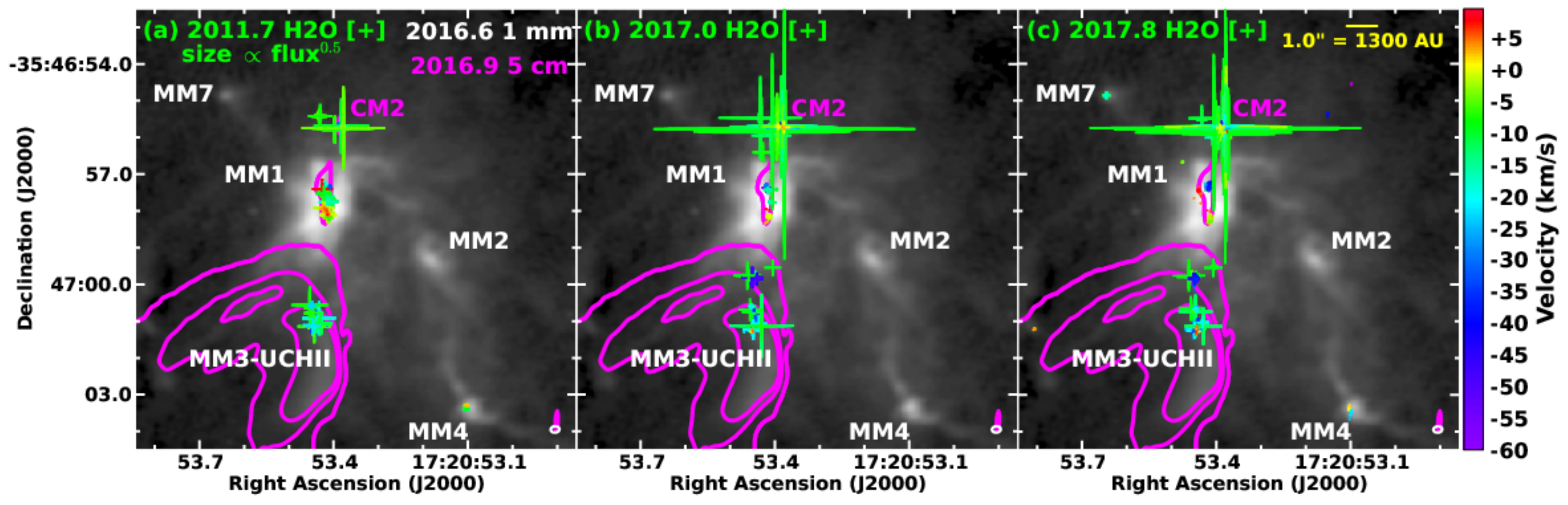}
\caption{(a) Water maser positions ($+$ symbols) from epoch 2011.7 \citep{Brogan16} are overlaid on an epoch 2016.6 ALMA 1~mm continuum image in greyscale \citep{Hunter17}. Epoch 2016.9 JVLA 5~cm continuum contours are shown in magenta with levels of $0.022\times [4,260,600]$~\mjb\/. The sizes of the maser symbols are proportional to (flux density)$^{0.5}$, while velocity is indicated by the 
color of the symbol. Continuum sources are labeled for reference; sources with millimeter emission are labeled in white. Not shown is the new maser position called I-South, which is located $30''$ south of the image center (see Table~\ref{waterassoc}). (b) and (c) Same as (a) but the epoch 2017.0 and epoch 2017.8 water masers are overlaid, respectively. The synthesized beams are shown in the lower right, with the same color as the wavelength label in panel (a).
\label{water}}
\end{figure*} 

\subsection{ALMA Observations}

The ALMA 1~mm observations presented in this paper are from 
project 2015.A.00022.T and have been described in earlier publications which showed the 1~mm dust continuum image, as well as the integrated intensity image 
from a methanol transition \citep{Hunter18}, and spectra that exhibited various emission lines of methoxymethanol \citep{McGuire17}. Due to the copious hot 
core line emission emanating from MM1 and MM2, extracting a list of relatively line free continuum channels to perform uv continuum subtraction for this 
source is challenging. 

To this end, we have developed a technique of identifying relatively line-free channels that first uses the corrected sigma 
clipping method (c-SCM) approach of STATCONT \citep{Sanchez18} to define the continuum level in each spectral window, then chooses a list of channels that 
best match that level over a representative ensemble of individual spatial pixels. These channels are then used to perform uv continuum subtraction in CASA.  In more detail, we 
first extract spectra from a sample of pixels ($\sim 10-20$) in the dirty line+continuum cube that represent the range of emission/absorption properties 
present in the field of view (spectral line peak positions, continuum peaks, and outflows, for example). For each spectrum, the c-SCM method is employed to 
find the sigma-clip continuum level (SCL) for that pixel using parameter $\alpha$=3. The SCL is then corrected for noise by $-0.5\sigma_{\rm SCM}$ (case~D of STATCONT). Second, we compute a spectrum of the absolute deviation from the SCL for each pixel.  We then form an aggregate deviation spectrum (ADS) from the maximum value per channel over all deviation spectra.  Finally, we define the line-free channels as those whose ADS value is below a specified level, which is adjusted manually until the line contamination for all the representative spectra is low, while retaining sufficient line-free channels spanning the window to enable successful uv-continuum subtraction. We note that while uv-continuum subtraction is sensitive to sampling across the whole spectral window, it is relatively insensitive to the total fraction of bandwidth included. For \ngci, the continuum bandwidth identified using this method is typically $10-20\%$ of the total observed spectral bandwidth. This method ensures that significant line emission is avoided across the range of spatio-spectral properties present, while still permitting uv-continuum subtraction, a method which provides distinct advantages when one needs to image lines in the presence of strong continuum emission \citep{Rupen99}. 

For this paper, we also present for the first time images of the CS (6-5) spectral line transition at 293.9120865~GHz \citep{Gottlieb03} that were imaged with a spectral resolution of 2~\kms.  These data are sensitive to smooth spatial structures $\lesssim 6\arcsec$ in size. The full width at half power of the primary beam of the ALMA 12~m antennas at this frequency is only $20''$; primary beam correction was applied to the images presented here. The synthesized beam of the 1~mm continuum image is $0\farcs24 \times 0\farcs17$ ($-84\arcdeg$), while the synthesized beam of the CS (6-5) image is $0\farcs28 \times 0\farcs21$ at $-84\arcdeg$. The sensitivity of the 1~mm continuum is 0.6~\mjb\/, while the sensitivity of the CS (6-5) channels free of significant emission is 3~\mjb. 

\begin{deluxetable*}{cccccc}  
\tabletypesize{\scriptsize}
\tablewidth{0pc}
\tablecaption{Fitted properties of the 22~GHz water masers in the epoch 2011.7 data \label{water2011table}}  
\tablecolumns{7}
\tablehead{\colhead{Number} &  \colhead{Association} & \colhead{Velocity channel} & \multicolumn{2}{c}{Fitted Position (J2000)} & \multicolumn{1}{c}{Flux Density} \\    
& & (\kms) & \colhead{R.A.} & \colhead{Dec.} &  (Jy)\tablenotemark{a} }
\startdata
  1 &   MM1-W3 & -64.50 & 17:20:53.4189 & -35:46:57.368 & 0.130 (0.018)\\
  2 &   MM1-W3 & -64.25 & 17:20:53.4189 & -35:46:57.356 & 0.203 (0.020)\\
  3 &   MM1-W3 & -64.00 & 17:20:53.4200 & -35:46:57.368 & 0.143 (0.018)\\
  4 & UCHII-W1 & -41.50 & 17:20:53.4461 & -35:47:01.083 & 0.347 (0.018)\\
  5 & UCHII-W1 & -41.25 & 17:20:53.4478 & -35:47:01.124 & 1.19 (0.02)\\
  6 & UCHII-W1 & -41.00 & 17:20:53.4485 & -35:47:01.135 & 3.78 (0.02)\\
  7 & UCHII-W1 & -40.75 & 17:20:53.4487 & -35:47:01.137 & 7.39 (0.03)\\
  8 & UCHII-W1 & -40.50 & 17:20:53.4486 & -35:47:01.136 & 8.56 (0.03)\\
  9 & UCHII-W1 & -40.25 & 17:20:53.4484 & -35:47:01.136 & 5.80 (0.02)\\
 10 & UCHII-W1 & -40.00 & 17:20:53.4476 & -35:47:01.136 & 2.69 (0.02)\\

\enddata
\tablenotetext{a}{Fitted flux density in this channel; uncertainties in the fitted value are listed in parentheses, and do not include the flux calibration uncertainty of 10\%.}
\tablenotetext{}{(This table is available in its entirety in a machine-readable form in the online journal. A portion is shown here for guidance regarding its form and content.)}
\end{deluxetable*}

\begin{deluxetable*}{cccccc}   
\tabletypesize{\scriptsize}
\tablewidth{0pc}
\tablecaption{Fitted properties of the 22~GHz water masers in the epoch 2017.0 data\label{water2017table}}  
\tablecolumns{7}
\tablehead{\colhead{Number} &  \colhead{Association} & \colhead{Velocity channel} & \multicolumn{2}{c}{Fitted Position (J2000)} & \multicolumn{1}{c}{Flux Density} \\    
& & (\kms) & \colhead{R.A.} & \colhead{Dec.} &  (Jy)\tablenotemark{a} } 
\startdata
  1 &   CM2-W1 & -61.75 & 17:20:53.4074 & -35:46:55.852 & 0.186 (0.016)\\
  2 &   CM2-W1 & -61.50 & 17:20:53.4080 & -35:46:55.844 & 0.174 (0.016)\\
  3 &   CM2-W1 & -61.25 & 17:20:53.4075 & -35:46:55.830 & 0.125 (0.015)\\
  4 &   MM1-W3 & -61.25 & 17:20:53.4193 & -35:46:57.356 & 0.100 (0.015)\\
  5 &   MM1-W3 & -61.00 & 17:20:53.4203 & -35:46:57.344 & 0.111 (0.013)\\
  6 &   MM1-W3 & -60.75 & 17:20:53.4195 & -35:46:57.356 & 0.138 (0.011)\\
  7 & UCHII-W3 & -60.75 & 17:20:53.4488 & -35:46:59.888 & 0.0876 (0.0108)\\
  8 &   MM1-W3 & -60.50 & 17:20:53.4193 & -35:46:57.373 & 0.153 (0.010)\\
  9 & UCHII-W3 & -60.50 & 17:20:53.4474 & -35:46:59.881 & 0.0793 (0.0098)\\
 10 &   MM1-W3 & -60.25 & 17:20:53.4193 & -35:46:57.353 & 0.205 (0.009)\\

\enddata
\tablenotetext{a}{Fitted flux density in this channel; uncertainties in the fitted value are listed in parentheses, and do not include the flux calibration uncertainty of 10\%.}
\tablenotetext{}{(This table is available in its entirety in a machine-readable form in the online journal. A portion is shown here for guidance regarding
its form and content.)}
\end{deluxetable*}

\begin{deluxetable*}{cccccc}   
\tabletypesize{\scriptsize}
\tablewidth{0pc}
\tablecaption{Fitted properties of the 22~GHz water masers in the epoch 2017.8 data\label{water2017btable}}  
\tablecolumns{7}
\tablehead{\colhead{Number} &  \colhead{Association} & \colhead{Velocity channel} & 
\multicolumn{2}{c}{Fitted Position (J2000)} & \multicolumn{1}{c}{Flux Density} \\    
 &  & (\kms) & \colhead{R.A.} & \colhead{Dec.} & (Jy)\tablenotemark{a}} 
\startdata
  1 &   CM2-W1 & -77.75 & 17:20:53.3847 & -35:46:55.762 & 0.120 (0.004)\\
  2 &   CM2-W1 & -77.50 & 17:20:53.3846 & -35:46:55.748 & 0.492 (0.004)\\
  3 &   CM2-W1 & -77.25 & 17:20:53.3846 & -35:46:55.752 & 0.873 (0.004)\\
  4 &   CM2-W1 & -77.00 & 17:20:53.3847 & -35:46:55.755 & 0.620 (0.004)\\
  5 &   CM2-W1 & -76.75 & 17:20:53.3843 & -35:46:55.742 & 0.234 (0.004)\\
  6 &   CM2-W1 & -76.50 & 17:20:53.3843 & -35:46:55.713 & 0.0993 (0.0038)\\
  7 &   CM2-W1 & -76.25 & 17:20:53.3848 & -35:46:55.738 & 0.0632 (0.0038)\\
  8 &   CM2-W1 & -76.00 & 17:20:53.3849 & -35:46:55.746 & 0.0387 (0.0033)\\
  9 &   CM2-W1 & -75.75 & 17:20:53.3846 & -35:46:55.619 & 0.0244 (0.0033)\\
 10 &   MM1-W3 & -71.00 & 17:20:53.4155 & -35:46:57.259 & 0.0179 (0.0035)\\

\enddata
\tablenotetext{a}{Fitted flux density in this channel; uncertainties in the fitted value are listed in parentheses, and do not include the flux calibration uncertainty of 10\%.}
\tablenotetext{}{(This table is available in its entirety in a machine-readable form in the online journal. A portion is shown here for guidance regarding
its form and content.)}
\end{deluxetable*}

\begin{deluxetable*}{lcccccccccc}   
\tabletypesize{\scriptsize}
\setlength{\tabcolsep}{0.05cm}
\tablecaption{Brightness summary of 22~GHz water maser associations\label{waterassoc}}  
\tablecolumns{10}
\tablehead{\colhead{Association} & \multicolumn{2}{c}{Mean position (J2000)\tablenotemark{a}} & \multicolumn{3}{c}{Integrated flux} &
\multicolumn{2}{c}{Integrated flux ratios} &
\multicolumn{3}{c}{Highest flux density\tablenotemark{b}} \\ 
&  &  & \multicolumn{3}{c}{(Jy~\kms)} & & & \multicolumn{3}{c}{(Jy)} \\
& \colhead{R.A.} & \colhead{Dec.} & 2011.7 & 2017.0 & 2017.8 & 
[2017.0/2011.7] & [2017.8/2011.7] & 2011.7 & 2017.0 & 2017.8}
\startdata
  I-South  &  17:20:53.182  &  -35:47:29.10  & ... &  2.57  &  0.881  & new & ...  & ... &  1.66  &  0.856 \\
   CM2-W1  &  17:20:53.380  &  -35:46:55.73  &  1790  &  11100  &  12000  & 6.2 & 6.7 &  398  &  3920  &  4230 \\
   CM2-W2  &  17:20:53.431  &  -35:46:55.42  &  55.4  &  399  &  39.0  & 7.2 & 0.70 &  34.8  &  403  &  42.0 \\
      CM2 Total  &  ...  &  ...  &  1840  &  11500  &  12000  & 6.3 & 6.5 &  ...  &  ...  &  ... \\
   MM1-W1  &  17:20:53.416  &  -35:46:58.03  &  94.5  &  7.84  &  5.79  & 0.083 & 0.061 &  26.9  &  2.77  &  2.27 \\
   MM1-W2  &  17:20:53.410  &  -35:46:57.73  &  53.4  &  1.25  & ... & 0.023 & vanished &  19.4  &  1.18  & ...\\
   MM1-W3  &  17:20:53.422  &  -35:46:57.44  &  63.7  &  22.5  &  12.3  & 0.35 & 0.19 &  25.3  &  5.33  &  2.56 \\
  MM1-W4  &  17:20:53.447  &  -35:46:57.28  & ... & ... &  1.60  & ... & new  & ... & ... &  1.28 \\
      MM1 Total  &  ...  &  ...  &  212  &  31.6  &  19.7  & 0.15 & 0.093 &  ...  &  ...  &  ... \\
      MM4  &  17:20:53.104  &  -35:47:03.36  &  4.73  & ... &  0.709  & vanished & 0.15 &  2.07  & ... &  0.295 \\
     MM7  &  17:20:53.645  &  -35:46:54.85  & ... & ... &  3.81  & ... & new  & ... & ... &  4.51 \\
  NWflow  &  17:20:53.107  &  -35:46:54.68  & ... & ... &  0.292  & ... & new  & ... & ... &  0.235 \\
 UCHII-W1  &  17:20:53.437  &  -35:47:01.04  &  701  &  755  &  366  & 1.1 & 0.52 &  66.5  &  237  &  83.3 \\
 UCHII-W2  &  17:20:53.451  &  -35:47:00.61  &  126  &  49.4  &  135  & 0.39 & 1.1 &  80.2  &  30.1  &  68.6 \\
 UCHII-W3  &  17:20:53.452  &  -35:46:59.81  & ... &  109  &  60.5  & new & ...  & ... &  46.5  &  60.5 \\
 UCHII-W4  &  17:20:53.405  &  -35:46:59.54  & ... &  7.68  &  10.4  & new & ...  & ... &  15.9  &  19.1 \\
UCHII-W5  &  17:20:53.806  &  -35:47:01.21  & ... & ... &  0.465  & ... & new  & ... & ... &  0.461 \\
    UCHII Total  &  ...  &  ...  &  827  &  921  &  572  & 1.1 & 0.69 &  ...  &  ...  &  ... \\
    Grand Total  &  ...  &  ...  &  2880  &  12500  &  12600  & 4.3 & 4.4 &  ...  &  ...  &  ... \\

\enddata
\tablenotetext{a}{This is the flux-weighted centroid of all spots in the epoch of first appearance. The differences in centroids between epochs are less than $0\farcs08$ with the following exceptions: UCHII-W1 and W2 which show bulk southward motion described in \S~\ref{bulk}; MM1-W1 which shows a large drop in total emission described in \S~\ref{mm1drop}; and MM4 which shows a new spatial component in 2017.8 (Fig.~\ref{MM4}). }
\tablenotetext{b}{These are the highest values for the respective components from the Flux Density column in Tables~\ref{water2011table}, \ref{water2017table}, and \ref{water2017btable}.}
\end{deluxetable*}

\begin{deluxetable*}{lccccc}   
\tabletypesize{\scriptsize}
\setlength{\tabcolsep}{0.2cm}
\tablecaption{Velocity summary of 22~GHz water maser associations\label{waterassoc2}}  
\tablecolumns{10}
\tablehead{\colhead{Association} & \multicolumn{3}{c}{LSR velocity: minimum, peak, maximum (\kms)\tablenotemark{a}} \\ 
& Epoch 2011.7 & Epoch 2017.0 & Epoch 2017.8}
\startdata
  I-South  & ... &  2.75, 4.00, 6.75  &  2.50, 3.75, 4.50 \\
   CM2-W1  &  -37.25, -5.25, 2.50  &  -61.75, -8.00, 4.50  &  -77.75, -7.75, 10.00 \\
   CM2-W2  &  -9.75, -9.00, -4.75  &  -10.50, -9.00, -8.00  &  -10.00, -9.25, -8.75 \\
      CM2 Total  &  -37.25, -5.25, 2.50  &  -61.75, -8.00, 4.50  &  -77.75, -7.75, 10.00 \\
   MM1-W1  &  -8.25, -2.00, 9.75  &  -3.75, 0.25, 9.00  &  -4.00, 0.25, 7.25 \\
   MM1-W2  &  -21.50, -20.25, -7.75  &  -15.00, -5.50, -5.50  & ...\\
   MM1-W3  &  -64.50, 8.75, 9.75  &  -61.25, -58.75, -9.00  &  -71.00, -26.25, -14.50 \\
  MM1-W4  & ... & ... &  -3.25, 8.25, 14.00 \\
      MM1 Total  &  -64.50, -2.00, 9.75  &  -61.25, -58.75, 9.00  &  -71.00, -26.25, 14.00 \\
      MM4  &  -13.75, -11.75, 3.50  & ... &  -23.25, -22.00, 2.75 \\
     MM7  & ... & ... &  -18.50, -10.25, -9.75 \\
  NWflow  & ... & ... &  -69.00, -60.75, -59.75 \\
 UCHII-W1  &  -41.50, -24.25, 1.50  &  -39.00, -11.50, 6.00  &  -45.00, -10.50, 10.75 \\
 UCHII-W2  &  -36.25, -9.25, -6.75  &  -40.25, -8.00, -7.50  &  -35.50, -11.75, -4.50 \\
 UCHII-W3  & ... &  -60.75, -7.50, -6.75  &  -69.00, -7.50, -2.75 \\
 UCHII-W4  & ... &  -8.00, -8.00, -7.75  &  -8.00, -8.00, -7.50 \\
UCHII-W5  & ... & ... &  3.25, 4.25, 5.00 \\
    UCHII Total  &  -41.50, -9.25, 1.50  &  -60.75, -11.50, 6.00  &  -69.00, -10.50, 10.75 \\
    Grand Total  &  -64.50, -5.25, 9.75  &  -61.75, -8.00, 9.00  &  -77.75, -7.75, 14.00 \\

\enddata
\tablenotetext{a}{Emission is not necessarily detected across the entire range from minimum to maximum velocity.}
\end{deluxetable*}

\begin{figure}[ht!]   
\includegraphics[width=1.0\linewidth]{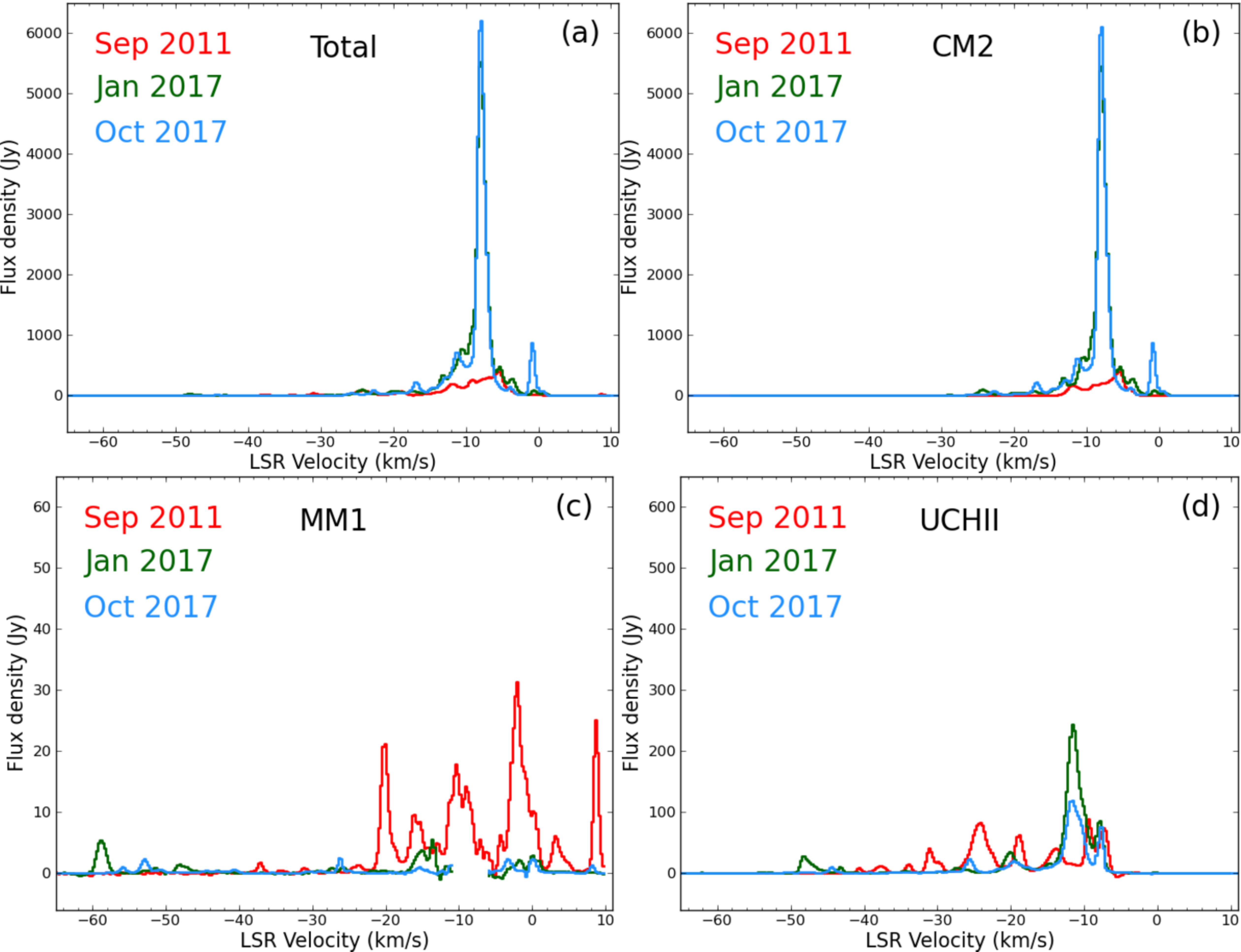}
\caption{Integrated spectra of the water maser emission from epoch 2011.7 (red line), 2017.0 (green line), and 2017.8 (blue line) constructed by summing the flux density from all directions showing emission: (a) summed over all spots; (b) CM2 only; (c) MM1 only; (d) the MM3-UCHII region only.  Note the factors of 100 and 10 change in the y-axis scale between panel b and panels c and d.  The intentional gap in the 2017 spectra of MM1 in panel c near the LSR velocity is to avoid confusion from the sidelobes of the strong feature in CM2 in those channels (see \S\ref{vlaobs}).
\label{h2o_spectra}}
\end{figure} 

\section{Results}

\begin{figure*}[h!]   
\includegraphics[width=1.0\linewidth]{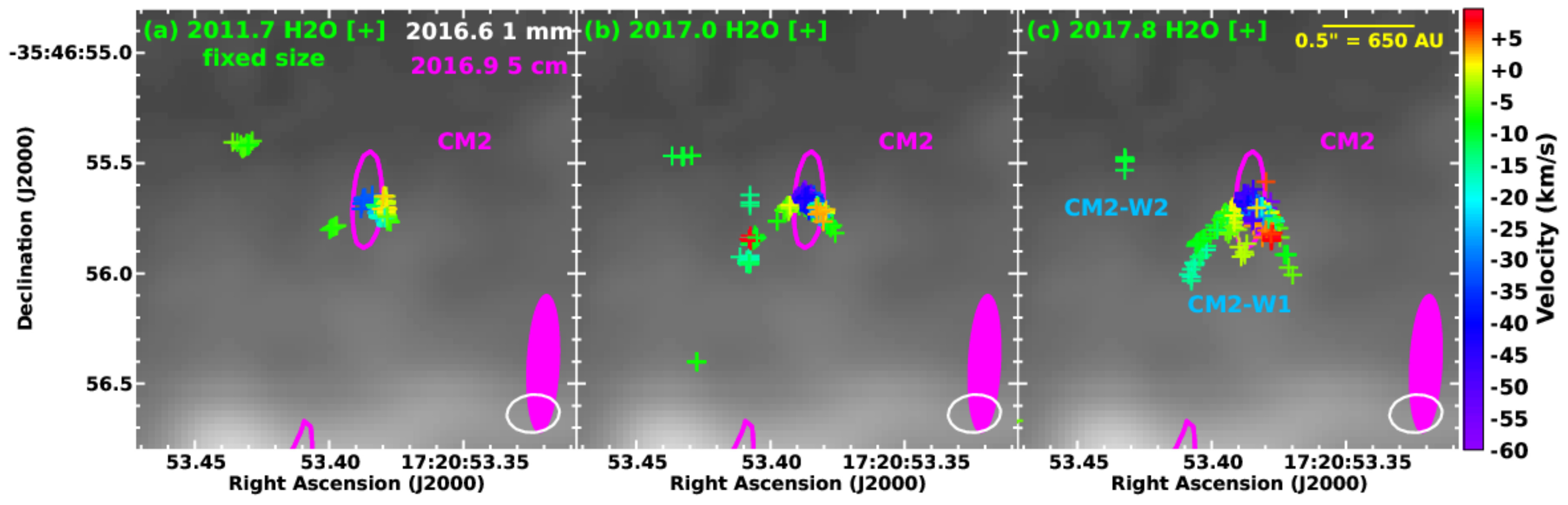}
\caption{Similar to Fig.~\ref{water}, but showing a close-up view of the water masers toward the cm-synchrotron source CM2 for the three epochs. VLA 5~cm continuum contours from epoch 2016.9 \citep{Hunter18} are shown in magenta at levels of $0.022\times [4]$ \mjb. In order to better see the morphology of the water masers, the spots are plotted on a fixed size scale. The synthesized beams are shown in the lower right, with the same color as the wavelength label in panel (a).
\label{CM2water}}
\end{figure*}

\begin{figure*}[ht!]   
\includegraphics[width=1.0\linewidth]{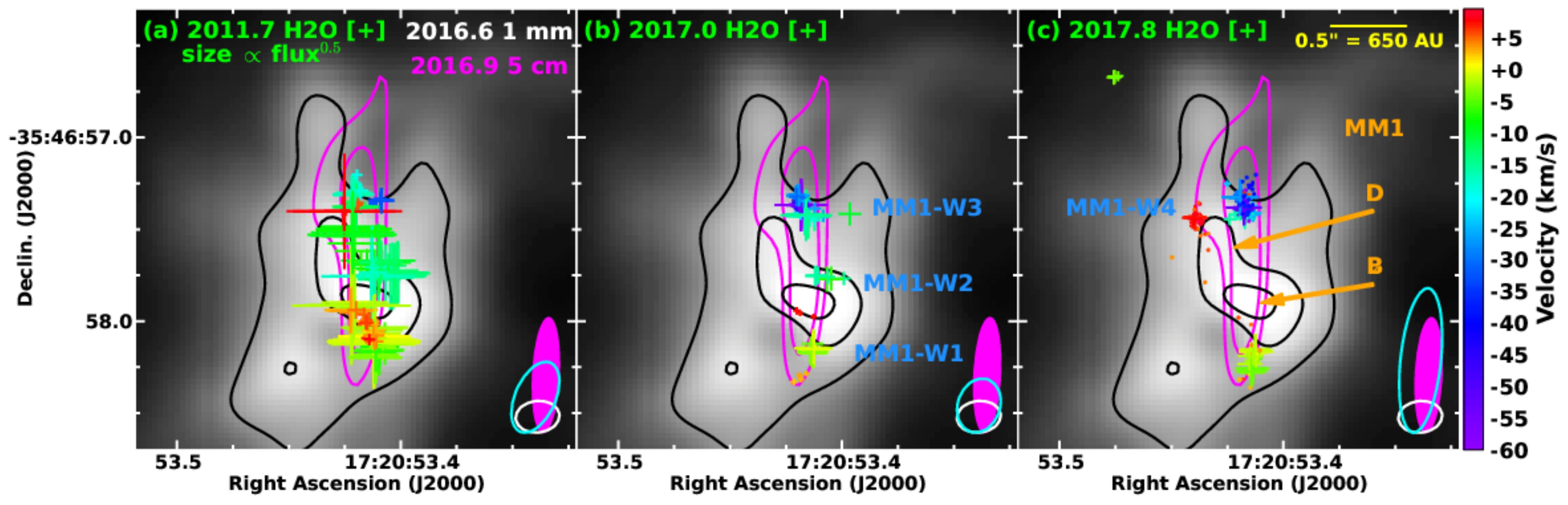}
\caption{Similar to Fig.~\ref{water}, except showing a smaller field of view toward MM1. VLA 5~cm continuum contours from epoch 2016.9 \citep{Hunter18} are shown in magenta at levels of $0.022\times [4, 9]$ \mjb. ALMA 1.0~mm data are shown in grey scale and black contours with levels of 
0.23, 0.43, and 0.53 \jb. The locations of the millimeter sources MM1B and MM1D are labeled for reference, and the synthesized beams are shown in the lower right corner in white, cyan, and magenta corresponding to 1.0~mm, 1.35~cm, and 5~cm. The synthesized beams are shown in the lower right, with the same color as the wavelength label in panel (a).
\label{waterzoom}}
\end{figure*}

\subsection{Water masers}

\begin{figure*}[ht!]   
\includegraphics[width=1.0\linewidth]{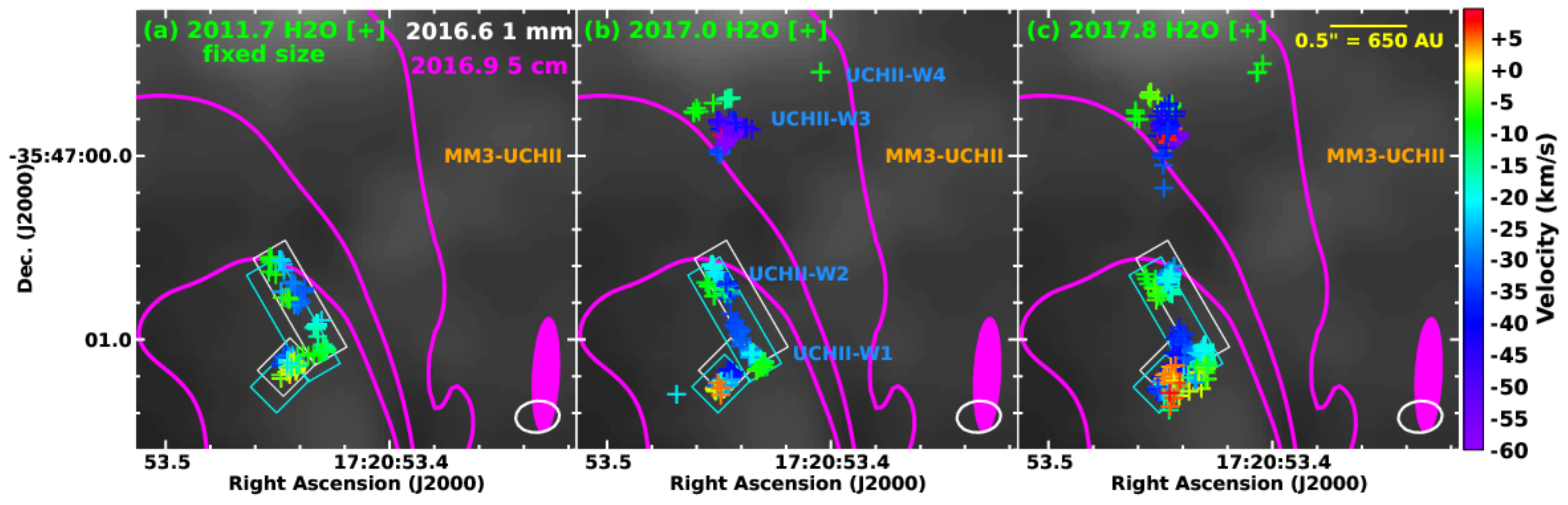}
\caption{Similar to Fig.~\ref{water}, except showing a smaller field of view toward the MM3-UCHII region masers. VLA 5~cm continuum contours from epoch 2016.9 \citep{Hunter18} are shown in magenta at levels of $0.022\times [4, 260,  600]$ \mjb. Each panel also shows two white and two cyan colored rectangular boxes toward the UCHII-W1 and W2 maser associations; the position of the boxes is the same in each panel (also see \S\ref{MM3sec} and \S\ref{bulk}). In order to better see the morphology of the water masers, the spots are plotted on a fixed size scale. The synthesized beams are shown in the lower right, with the same color as the wavelength label in panel (a).
\label{MM3}}
\end{figure*} 

We fit each channel of the water maser cube that had significant emission ($\ge 6\sigma$) with an appropriate number of Gaussian point sources, using the pixel position of each emission peak as the initial guess for the fitted parameters. A few weak sources that could be attributed to imaging artifacts in the dynamic-range limited channels were discarded despite otherwise meeting the detection threshold. A few of the most complicated channels exhibiting many closely-spaced components required a manual iterative procedure to fix the declination position of one or two components in order for the simultaneous fit for all components to converge. The fitted positions and flux densities of the water masers are given in Tables~\ref{water2011table}, \ref{water2017table}, and \ref{water2017btable} for the 2011.7, 2017.0 and 2017.8 data, respectively.  The fitted strength of the emission was corrected for primary beam attenuation. 

The locations of the fitted maser positions are shown in Figure~\ref{water}.  To summarize the different maser locations in a compact format, we have identified 
maser ``groups'' toward the centimeter and millimeter continuum sources that had water masers detected toward them: MM1, CM2, MM3-UCHII, MM4, MM7, and a few outliers called NWflow and I-South. When the water masers within these groups are concentrated into spatially distinct ``associations'' they are further labeled with the suffix -W1, -W2, etc for ease of referral.  The naming of these associations parallels that used for the 6.7~GHz methanol and 6.035~GHz OH masers reported in \citet{Hunter18} toward this source. The assignment of each fitted maser spot to an association is provided in a column in Tables~\ref{water2011table}, \ref{water2017table}, and \ref{water2017btable}.
These data show that the masers have flared significantly toward the 5~cm synchrotron source CM2, and that the broad range of maser velocity components observed previously persists in the current epochs \citep[see Fig.~\ref{water}, as well as][]{Brogan16,Titmarsh16,Breen10,FC99}. 
The change in the total maser emission and the changes in the individual associations are summarized by the integrated spectra in Figure~\ref{h2o_spectra} and in the summary Tables~\ref{waterassoc} and \ref{waterassoc2}.  We next discuss the individual groups (and associations within them) in more detail.

\subsubsection{CM2}

In all three VLA epochs presented here, the masers toward and close to the 5~cm point source CM2 are the strongest in the region (Fig.~\ref{water},~\ref{CM2water}; CM2-W1 and CM2-W2).  Between mid-2011 and 2017, they exhibited a large increase in peak flux density, with the peak increasing from $<156$~Jy in 2011.4 \citep{Titmarsh16} and 398~Jy in 2011.7 \citep{Brogan16} to 3910~Jy in 2017.0 and 4208~Jy in 2017.8.  The peak intensity in 2011.7 is similar to the value of 246~Jy observed at a comparable velocity in the first VLA observations of this region with 1.32~\kms\/ channels \citep[epoch 1984.5,][]{FC99}, which further extends the baseline of the pre-outburst emission from CM2 to three decades. Figure~\ref{CM2water} shows a close-up view of the CM2 water masers (with fixed symbol size) for the three epochs. This figure demonstrates that especially during the 2017.0 and 2017.8 epochs, the water masers coincident with CM2-W1 trace out a distinctive bow shock morphology pointing back toward MM1B, with the most-blueshifted masers forming the apex of the bow shock. The nature of this bow shock is discussed further in \S \ref{CM2disc}.

\subsubsection{MM1}
In MM1, we label the initial (epoch 2011.7) associations of masers as follows: the association toward and south of MM1B as MM1-W1, the association between MM1B and MM1D as MM1-W2, and the association north of MM1D as MM1-W3. Comparison of the 2011.7 and 2017.0 epochs reveals that all the 2011.7 maser sites have significantly weakened.  Most notably, MM1-W2 disappears entirely by the 2017.8 epoch, see Fig.~\ref{waterzoom}.  In contrast, new blueshifted features have appeared within MM1-W3 (to the northwest of MM1D) in the 2017.0 epoch, persisting into 2017.8.  Also, a new association of redshifted features has appeared northeast and east of MM1D in epoch 2017.8, which we label MM1-W4.  The peak feature of MM1-W4 is 1.28~Jy at 8.25~\kms, which would have been a $100\sigma$ detection had it been present in that channel in the 2017.0 data.  Interestingly, the MM1-W4 masers coincide with the slight eastward extension of the 5~cm emission that lies just north of MM1D.

\subsubsection{MM3-UCHII Region}\label{MM3sec}

As shown in Figure~\ref{MM3}, in the first epoch (2011.7) only two maser associations were detected toward MM3-UCHII: UCHII-W1 and UCHII-W2, each forming a roughly linear structure, in roughly perpendicular directions. These associations persisted to the 2017.0 and 2017.8 epochs, though in the 2017.8 epoch they have lost much of their ordered morphology. The two later epochs also show two new maser associations further north but still coincident with the UCHII region: UCHII-W3 and UCHII-W4. Of particular note is the fact that the UCHII-W1 and UCHII-W2 associations appear to have undergone bulk motion toward the south-southeast. This is demonstrated by the white boxes in Figure~\ref{MM3}, which encompass these associations in the 2011.7 epoch; in contrast, the cyan boxes encompass these associations in the 2017.0 epoch data.  The offset between the colored boxes is the  proposed bulk motion. The more chaotic morphology of these associations by 2017.8 precludes a definitive judgment, but their locations in this epoch are also in general agreement with the bulk motion interpretation. This idea is explored further in \S\ref{bulk}. 

\begin{figure}[h!]   
\includegraphics[width=1.0\linewidth]{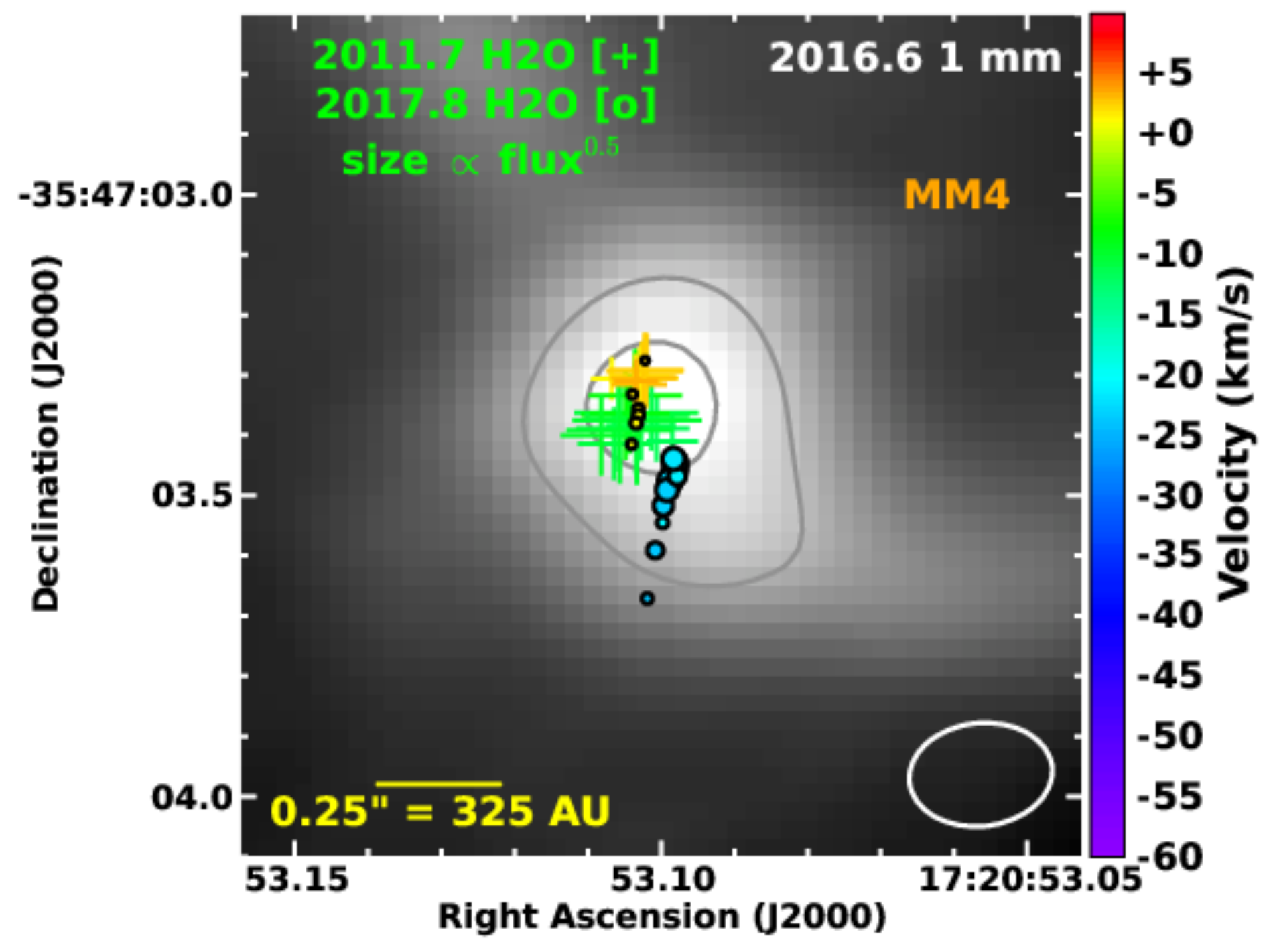}
\caption{Close-up view of the water masers detected towards MM4 in the 2011.7 ($+$ symbols) and 2017.8 (filled circles) epochs. The ALMA 1.0~mm image is shown in grey scale and grey contours with levels of 0.07 and 0.14 \jb\/. The synthesized beam of the 1~mm image is shown in the lower right.
\label{MM4}}
\end{figure} 

\subsubsection{MM4}
In both the 2011.7 and 2017.8 epochs, weak water masers are detected near the MM4 continuum peak (Fig.~\ref{MM4}). In the more recent epoch, the masers form two linear features that are roughly perpendicular to the bipolar outflow direction described in \S\ref{outflow}. The two linear features are offset from each other in RA by $\sim$100 au; the more easterly feature is slightly redshifted, while the western feature is blueshifted. Only the eastern feature was detected in the 2011.7 epoch, and its masers were significantly stronger then, by a factor of almost seven compared to the 2017.8 epoch (see Table~\ref{waterassoc}). The non-detection of masers in the MM4 group in the 2017.0 epoch must be due to variability, as the strongest 2011.7 epoch maser 
(0.72~\jb\/ at +2.25~\kms) would have been a $21.6\sigma$ detection in the 2017.0 epoch. In 2017.8, masers in the eastern feature are detected at $6-11~\sigma$. 


\begin{figure}[h!]   
\includegraphics[width=1.0\linewidth]{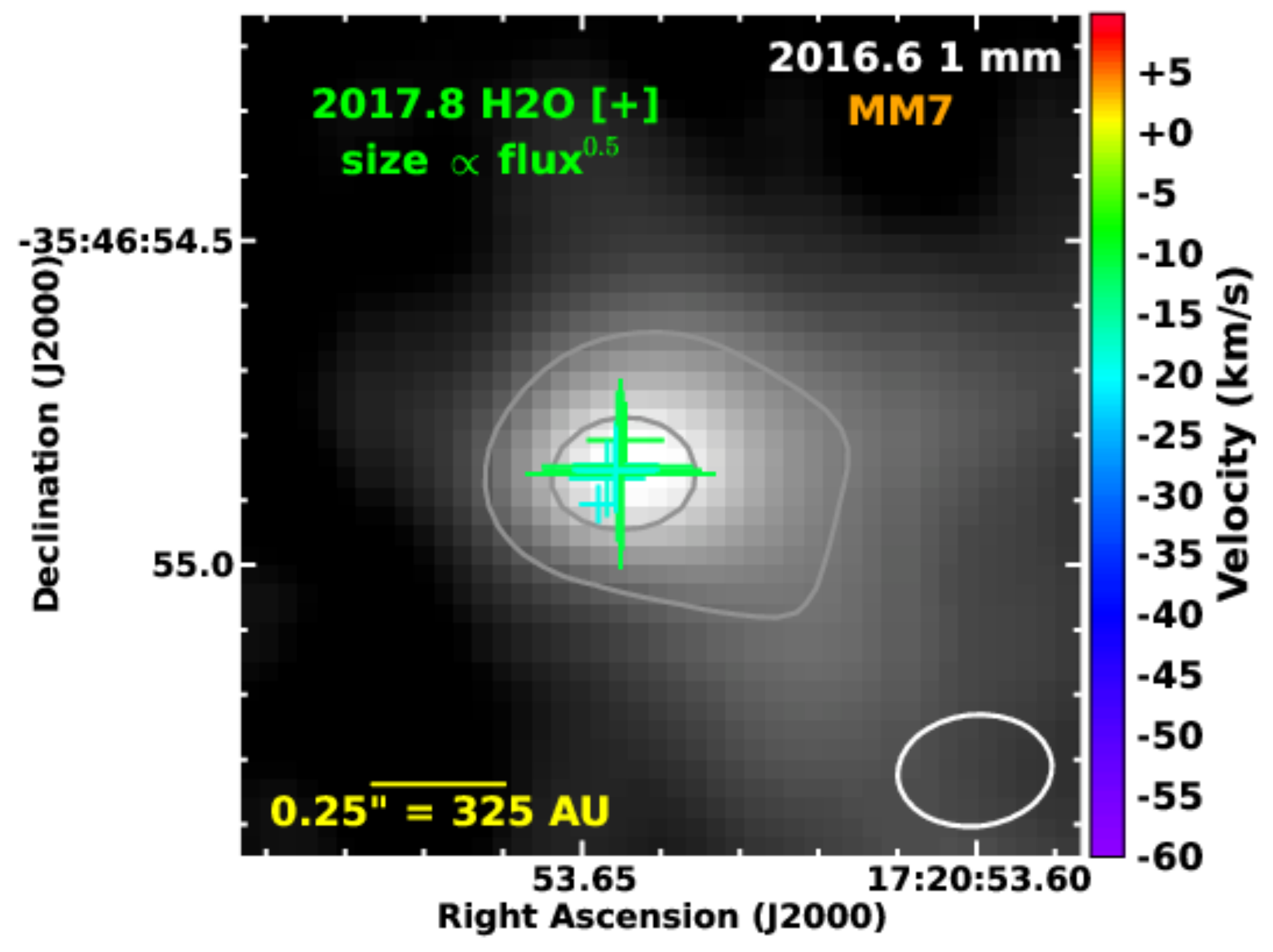}
\caption{Close-up view of the water masers detected towards MM7 in the 2017.8 epoch ($+$ symbols). The ALMA 1.0~mm image is shown in grey scale and grey contours with levels of 0.01 and 0.03 \jb\/. The synthesized beam of the 1~mm image is shown in the lower right.
\label{MM7}}
\end{figure} 

\subsubsection{MM7}  
In the 2017.8 epoch, water masers are detected for the first time toward the millimeter core MM7, with a peak intensity of 4.51~\jb\/ at -10.25~\kms\/ (Figure~\ref{MM7}). The $3\sigma$ upper limits in this channel for the earlier epochs are 0.54 and 0.87~\jb, respectively.  Considering all channels with detections in 2017.8, the highest intensity increase factors relative to epochs 2011.7 and 2017.0 are $>9.3$ (at -17.50~\kms) and $>11.9$ (at -17.75~\kms), respectively.

\subsubsection{NWflow}
Two additional new detections of water masers in the 2017.8 epoch lie along an axis extending northwest of MM1 with peak flux densities of 0.045~Jy and 0.235~Jy.  Because they align with a collimated feature in the CS images (see \S~\ref{outflow}), we label them NWflow. The $3\sigma$ upper limits in their channels of peak emission ($-68.5$ and $-60.75$~\kms, respectively) are 0.03~\jb\/ in epoch 2017.0 and 0.07~\jb\/ in epoch 2011.7.

\subsubsection{\ngci-South}
Finally, in the 2017.0 and 2017.8 epoch data we detected a new water maser located about $30''$ south of the central protocluster, emitting at redshifted velocities from +2.75 to +7.0~\kms\/ with a peak flux density of 1.66 and 0.856~Jy, respectively, in the two epochs. \added{Because this maser lies well outside the fields of view of the figures in this paper,} we call it \ngci-South in order to distinguish it from the masers in the central protocluster (see Table~\ref{waterassoc}).  In the peak channel (4.0~\kms), it was not detected in the 2011.7 epoch data with a $3\sigma$ upper limit of 0.065~\jb, leading to a lower limit of 25 for the increase factor between 2011.7 and 2017.0.
 
\subsection{Other maser transitions}
A few \methanol\/ maser lines fall within our coarse spectral resolution windows, such as the Class~II 2(1)--3(0)~E transition at 19.9674~GHz \citep{Menten89,Ellingsen04}.  We imaged the corresponding channel of our coarse resolution data and found that the positions of the emission, toward MM2 and the MM3-UCHII~region, agree with the VLA positions reported by \citet{Brogan16}.  No other new positions were detected.

\begin{figure*}[ht!]   
\includegraphics[width=1.0\linewidth]{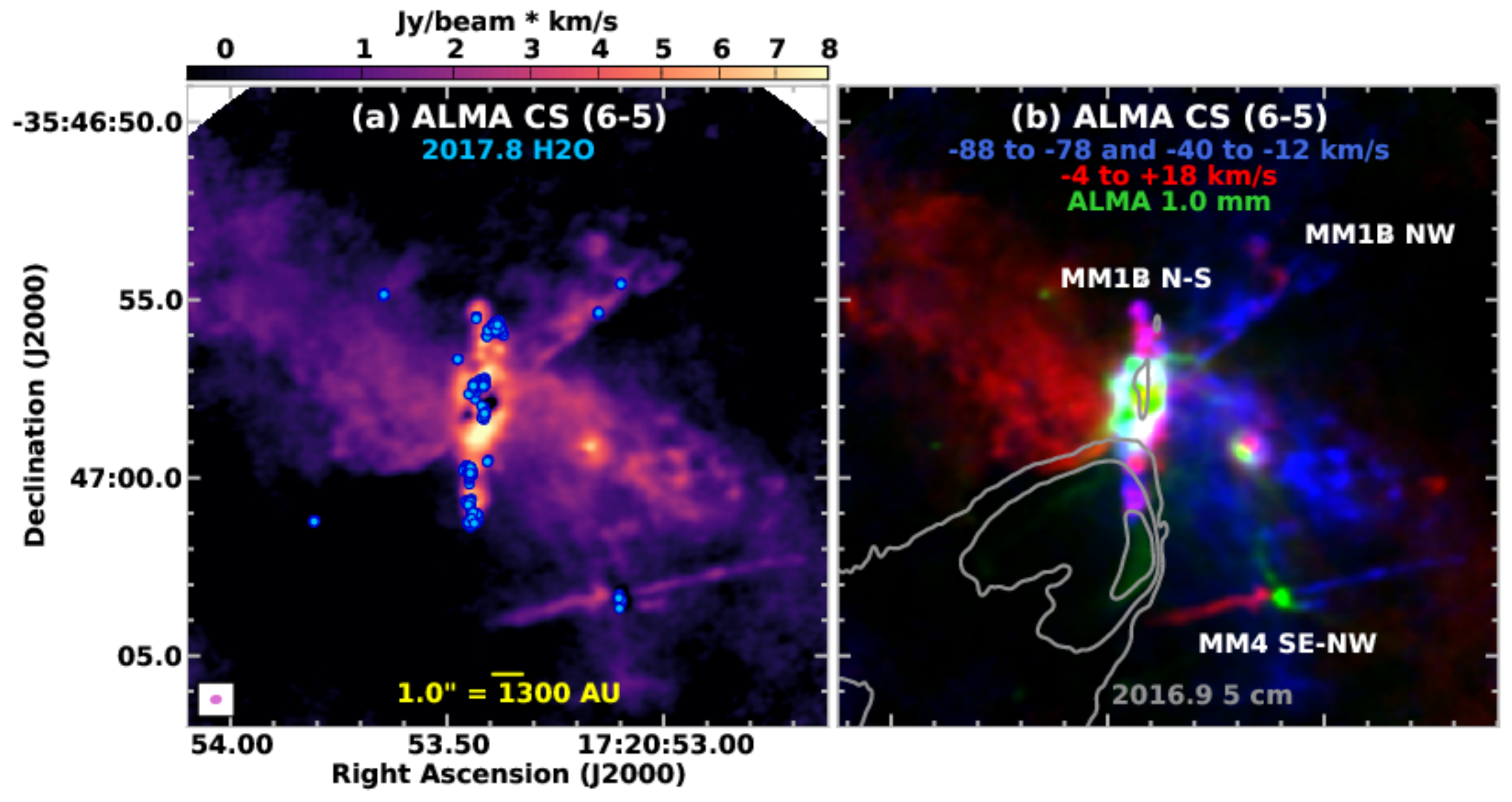}
\caption{The integrated intensity of ALMA CS(6-5) line emission is shown in both panels, with a synthesized beam of $0\farcs28 \times 0\farcs21$ at $-84\arcdeg$.
In (a) the emission has been integrated from $-88$ to $-78$ \kms\/, $-40$ to $-12$ \kms\/, and $-4$ to $+40$ \kms\/; the systemic velocity is $\sim -7$~\kms\/ (see \S~\ref{outflow} for description of velocity ranges). The positions of the 2017.8 epoch water masers are overlaid as blue circles, and the CS (6-5) synthesized beam is shown in the lower left. In (b) the integrated emission from $-88$ to $-78$ and $-40$ to $-12$ \kms\/ is color-coded blue and the integrated emission from -4 to +40 \kms\/ is color-coded red. The 1.0~mm dust continuum emission is shown in green (see Fig.~\ref{water} for continuum source names). The three newly detected outflows are labeled in white. In addition, the epoch 2016.9 5 cm continuum is shown in grey contours with levels of $0.022\times [4, 160, 800]$~\mjb\/, to show the location and morphology of the MM3-UCHII region, as well as the jet associated with MM1B, and the synchrotron source CM2.
\label{CS}}
\end{figure*} 

\subsection{Outflows detected in CS (6-5)}\label{outflow}

The CS (6-5) transition is well suited to trace moderately dense, warm gas, with an upper state energy $E_{up}=49.4$~K and a critical density of $n_{crit}=9\times 10^6$~\cc\/ using
\begin{equation}
n_{cr} = \frac{A_{ul}}{\gamma _{ul}},
\end{equation}
where $A_{ul}$ is the Einstein $A$ coefficient of the transition (5.23~$\times$~10$^{-4}$~s$^{-1}$) and $\gamma _{ul}$ is the collisional coefficient at 200 K \citep[$6.02~\times~10^{-11}$~cm$^3$~s$^{-1}$,][]{Schoier05,Lique06}.
Figure~\ref{CS}a shows the integrated intensity of CS (6-5) emission toward \ngci\/ including the velocity ranges $-88$ to $-78$ \kms\/, $-40$ to $-12$ \kms\/, and $-4$ to $+40$ \kms\/ (these ranges exclude channels near the systemic velocity of $\sim -7$~\kms\/, and channels with significant confusing hot core line emission toward MM1). Figure~\ref{CS}b shows the same integrated intensity velocity ranges, but color-coded red and blue. From these images it is clear that several distinct collimated outflow structures are detected in the CS (6-5) transition. 

On the largest size scale, the inner portion of the large scale NE-SW outflow, well known from previous single dish studies \citep[][]{Qiu11,Leurini06,McCutcheon00}, is readily apparent in these data, though its full projected extent ($\sim 50\arcsec$) lies well beyond the primary beam of these single pointing ALMA data (FWHM $20\arcsec$). The origin of this outflow has long been debated because at lower angular resolution it could reasonably emanate from either a protostar in MM1 or MM2 \citep[e.g.,][]{Beuther08}. These new high resolution and fidelity ALMA data reveal that indeed the majority of emission arises from one of the MM1 protostars, especially the red-shifted NE lobe, but there is also a one-sided blueshifted component from MM2 that extends to the SW coincident with the larger blueshifted SW lobe. \citet{Qiu11} estimate a dynamical 
age for the large NE-SW outflow of $2.6\times 10^3$~yr.

A number of other smaller, highly collimated outflows are evident from these data for the first time. The brightest emanates from MM1B in a roughly North-South direction (PA=$-6\arcdeg$) with a projected linear extent of $\sim 6.2\arcsec$ (8060 au), and velocity extent from $-40$ to $+18$ \kms\/. This outflow will hereafter be called the MM1B N-S outflow. The majority of water masers in the \ngci\/ region are coincident with bright knots of CS (6-5) emission in this outflow (see Fig.~\ref{CS}a). Since it also has the same N-S orientation as the 5~cm jet-like emission centered on MM1B \citep[see Fig.~\ref{waterzoom}, \ref{CS}b, and][]{Brogan16}, we suggest the 5~cm jet arises from the ionized base of the MM1B N-S outflow. Interestingly, toward the northern lobe, a bright CS (6-5) knot lies just south of the CM2 water maser bow shock shown in Fig.~\ref{CM2water}. As demonstrated in Fig.~\ref{CS}b, the red- and blueshifted emission from the MM1B N-S outflow are superposed along the line of sight as expected for an outflow that lies nearly in the plane of the sky. For a maximum velocity extent of $33$~\kms\/ (from systemic) and a symmetric single-lobe extent of $3.1\arcsec$ we estimate the dynamical time is $\sim 580\,\tan(\theta_{inc})$~yr, where $\theta_{inc}$ is the inclination angle between the outflow axis and the plane of the sky. For $\theta_{inc}=10\arcdeg$, the dynamical time for the MM1B N-S outflow is only 102 yr. Interestingly, in Band 10 ALMA data we have also recently detected this outflow in CS (18-17)  ($E_{up}=402$~K, $n_{crit}=2\times 10^8$~\cc\/, and $\nu=880.9$~GHz) and the ground state transition of HDO ($1_{0,0}-0_{0,0}$) at $\nu=893.6$~GHz, i.e. thermal water emission \citep[][]{McGuire18}. The strong detection of CS (18-17) suggests the physical conditions in this outflow are warm and dense given the high energy and critical density of this transition, in agreement with past {\em Herschel} studies of high $J$ molecular line transitions toward protostellar outflows \citep[see for example][]{Goicoechea2015}. However, since the observed spatial scales of the Band 10 data are not well-matched to the CS (6-5) data we have not attempted to model the CS line emission.

A third, predominantly blueshifted, outflow lobe is detected extending to the NW of the MM1 region for $\sim 6.3\arcsec$ (8100 au), over a velocity extent of $-94$ to $+2$~\kms\/ at a position angle of $+138\arcdeg$ (Fig.~\ref{CS}).  Note that while the full velocity range of this outflow is not shown in Fig.~\ref{CS} (in order to avoid strong confusing hot core lines toward MM1), its overall morphology is well-represented. The APEX single dish observations of \citet{Qiu11} also detected high velocity blueshifted emission in CO (9-8) toward this area, albeit with an angular resolution of $6.4\arcsec$, such that the emission was unresolved. The opposing side of this outflow is not obvious in the CS (6-5) data, possibly because it is confused with the redshifted emission from the large-scale NE-SW flow. 
This outflow also appears to point back toward MM1B (so it is henceforth called MM1B NW), and two water masers (NWflow) are detected toward it in the 2017.8 epoch data (Fig.~\ref{CS}a). An isolated water maser $\sim 5\arcsec$ southeast of MM1 (UCHII-W5), with a redshifted velocity of $+4.25$~\kms\/, falls along a line joining MM1B and the NWflow water masers, suggesting that perhaps this maser traces the opposing side of the outflow, rather than the edge of the MM3-UCHII region, or perhaps that this location marks an interaction between the two structures. From the radio recombination line velocity of the MM3-UCHII region \citep[$-5$ to $-7$ \kms\/,][]{Hunter18}, it is plausibly co-distant with MM1. We estimate a dynamical time for the MM1B NW outflow of $\sim 450\,\tan(\theta_{inc})$~yr.

Finally, in CS (6-5) we have discovered a highly collimated bipolar SE-NW outflow emanating from MM4 (MM4 SE-NW) at a position angle of $+102\arcdeg$ (see Fig.~\ref{CS}). The velocity extent of this outflow is $-28$ to $+18$ \kms\/. The full (projected) linear extent is $\sim 9.1\arcsec$ (11830 au), however, the blueshifted emission extends $5.2\arcsec$, somewhat further than the red side. The smaller extent of the 
redshifted lobe may be due to interaction with the 
MM3-UCHII region if they are co-distant (which is currently unknown). Using the projected length of the blueshifted emission and maximum velocity extent from systemic of 25 \kms\/ we estimate a dynamical time of $\sim 1280\,\tan(\theta_{inc})$~yr. This outflow would have been spatially confused with the large-scale MM1 NE-SW bipolar outflow in past lower resolution molecular line data. 

\begin{figure*}[h!]   
\includegraphics[width=1.0\linewidth]{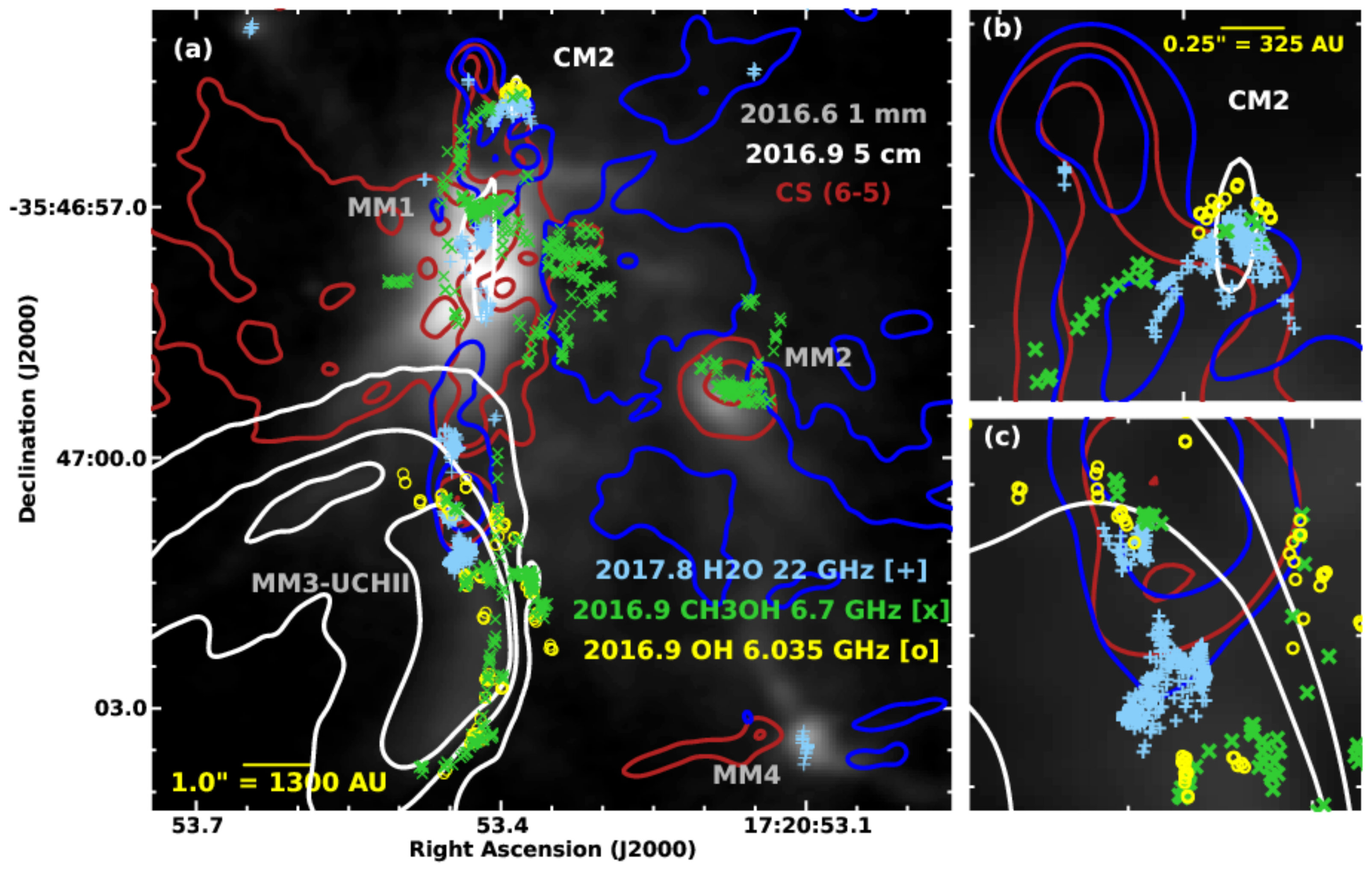}
\caption{Panel (a) shows the inner $9.6\arcsec$ of the \ngci\/ region in 1~mm continuum emission (greyscale) with blue and red contours overlaid showing the CS (6-5) integrated intensity emission with levels of [0.45, 1.1] and [0.55,1.1]~\jb*\kms\/, respectively. The blueshifted velocity range used for this image is $-24$ to $-12$~\kms\/, while the redshifted emission is integrated from $-4$ to $+8$~\kms\/; in both cases the velocity ranges were chosen to minimize contamination from hot core line emission from MM1 and MM2 while showing the key features of the MM1 N-S outflow. White epoch 2016.9 5~cm continuum contours are also shown at $0.022\times [4,260,600]$~\mjb\/. Superposed are light blue $+$ symbols showing the locations of the 2017.8 epoch water masers, as well as green $\times$ symbols and yellow $\circ$ symbols showing 2016.9 epoch CH$_3$OH 6.7~GHz masers and OH 6.035 GHz masers, respectively. Panels (b) and (c) show close-up views of the northern and southern ends of the N-S outflow where all three maser species are detected in close proximity; each FOV is $1.3\arcsec$.
\label{allmasers}}
\end{figure*} 

\section{Discussion}\label{disc}

In CS (6-5) we have discovered a new dynamically young outflow emanating from MM1B, toward which most of the water masers detected in the \ngci\/ region are found. 
In the following sections we present comparisons of how the water maser morphology and flaring compare to the 2016.9 epoch 6.7~GHz methanol and 6.035~GHz excited OH maser positions reported by \citet{Hunter18}. We also discuss the implied physical conditions and nature of the water maser emission.

\subsection{Comparison between post-outburst water, 6.7~GHz methanol and 6.035~GHz excited OH masers}

Figure~\ref{allmasers}a shows the locations of the 2017.8 epoch water masers, as well as the 2016.9 epoch 6.7~GHz methanol and 6.035~GHz excited OH masers reported by \citet{Hunter18} with respect to the 1~mm dust continuum, 5~cm emission, and the N-S outflow traced by CS (6-5). The north-south morphology of the ensemble of maser species is striking. Figures~\ref{allmasers}b and c show close up views of the two regions where the three maser species are in close (projected) proximity to each other -- the northern and southern extremes of the CS (6-5) N-S outflow. Additional information about these relationships is given in the following sections.

\subsection{Nature of the water maser flare in CM2 \label{CM2disc}}

\subsubsection{The water maser bow shock}

In star-forming regions, the 22~GHz water transition (between levels 644~K above ground) can arise because the upper level is a ``backbone'' level, which is efficiently coupled radiatively with other backbone levels \citep{deJong73}, in contrast to the lower level, which can de-excite via spontaneous emission, leading to an inversion \citep{Gray16}.  The upper level is pumped via collisions with H$_2$ in the warm (300-400~K), dense (10$^8$-10$^9$~\ccc) post-shock gas of shocks driven by protostellar jets and outflows, either dissociative $J$-shocks \citep{Hollenbach13} or non-dissociative magnetohydrodynamic $C$-shocks \citep{Kaufman96}. At high gas densities ($>10^{10}$~\ccc), the inversion is suppressed by collisions \citep{Hollenbach13}. 

The highly-blueshifted velocities seen near the apex of the spectacular bow shock pattern formed by the CM2-W1 water maser association toward the northern lobe of the MM1B N-S outflow (Figs.~\ref{CM2water},\ref{allmasers}), combined with the high proper motions measured for the masers in the southern lobe (\S~\ref{bulk}), gives strong evidence for the excitation of these water masers behind high-velocity dissociative shocks \citep{Hollenbach13}.
A general feature of bow shock models has the shocked gas moving away from the working surface and along a thin shell back toward the driving source \citep[e.g.][]{Blondin90,Taylor92,Norman82}.  This shell structure provides the required velocity-coherent path length that supports maser amplification along the limbs to the line-of-sight, which explains why most of the masers along the shell are near the systemic velocity of the source. Numerical simulations of jet-driven protostellar outflows \citep{Ostriker01} indicate that the radial transverse velocity of material being ejected from the working surface is of order the sound speed in $10^4$~K gas ($\sim$10~\kms\/).  As described by \citet{Gomez99} in the context of the protostellar jet in GGD~34, the expected backflow velocity can be a few times higher ($\sim$30~\kms) depending on the jet velocity and the degree of turbulence.  It is interesting that we see strongly blueshifted masers ($\sim 35$~\kms\/) relative to the systemic velocity ($\sim -7$~\kms) right at the apex of the bow shock, suggesting that these masers are tracing the origin of the backflow within the tangent plane to the bow shock. The lack of highly redshifted features along the same line of sight may indicate that the continuum emission being amplified by the masers at the tip of the shock is very compact in all three dimensions.

In the context of 3D bow shock models, fitting a parabola to the distribution of CM2-W1 water maser spots in the 2017.8 epoch yields a radius of curvature $r_0=0\farcs12 \approx 150$~au \citep[see][]{Gustafsson10}.  The strongest water masers occur on the western side of the apex at a projected distance of $\approx r_0$ from both the apex and the axis of the bow shock. The total width of the bow shock traced by CM2-W1 ($\sim$500~au) is similar to the size scale of maser arcs that are perpendicular to the outflow direction as seen in VLBI observations of more distant massive star formation regions like W49N \citep{Gwinn94}, and somewhat larger than the maser arcs recently resolved in AFGL~5142 \citep{Burns16} and IRAS~20231+3440 \citep{Ogbodo17}.  Thus, these maser features appear to be a common result of the impact of high-velocity jets from massive protostars.

\subsubsection{Excited OH and \methanol\/ masers downstream of the bow shock}

The bow shock pattern of the water masers in CM2-W1 also manifests in the excited-state OH masers reported by  \citep[CM2-OH1][]{Hunter18} but shifted by $\sim$120~au downstream (i.e. north; see Fig.~\ref{allmasers}b).  Taken alone, this observation would seem to favor a non-dissociative low-velocity shock because in this case nearly all of the free oxygen is incorporated into water \citep{Kaufman96}, leading to a low abundance of OH, in contrast to dissociative high-velocity shocks where the abundance of these molecules 
is more comparable \citep{Neufeld89}.  However, radiative transfer modeling of these OH transitions predicts that H$_2$ densities of $5 \times 10^7$~\ccc\/ are required to effectively pump these masers \citep{Etoka12,Cragg02}. Furthermore, these masers are efficiently pumped only at fairly low kinetic (gas) temperatures of 25 to 70~K \citep[see for example][]{Cragg02}. Thus, compared to water masers, these OH masers trace less dense, cooler gas, which is consistent with their downstream location where the gas is not yet fully compressed by the approaching shock.  We therefore conclude that the data are still consistent with a dissociative shock.

We also detect a few 6.7~GHz Class~II methanol masers downstream from CM2-W1 but upstream of the OH masers. These masers are pumped by infrared radiation and quench at densities above about $10^8$~\ccc\/ \citep{Cragg05}; thus they can survive in denser gas than OH, but will only appear in locations where the radiation received, possibly from the working surface itself \citep{Ostriker01}, is sufficiently strong.  Thus, their confinement to locations between OH masers and the water maser bow shock is not surprising.  The pumping of the methanol masers to the southeast of CM2-W1 (Fig.~\ref{allmasers}b) is most likely supported by the increased infrared radiation from MM1B itself, as it travels up the outflow cavity and outward through a few low extinction pathways as postulated by \citet{Hunter18}.

\subsubsection{The maser/outflow connection and synchrotron nature of CM2}

The alignment of CM2 and the water maser bow shock CM2-W1 with the orientation of the 5~cm jet arising from MM1B suggests that the two phenomena are associated. The north/south morphology of the water masers associated with MM1 and CM2 was first noted as a jet-like feature by \citet{Breen10} in reference to the original 1984 VLA C-configuration data of \citet{FC99}.  Our
detection of the MM1B N-S outflow in CS confirms the association of the masers with a bipolar outflow cavity.  Apparently, this previously-excavated cavity allowed the radiative energy of the current outburst (epoch $\sim 2015$) to propagate freely until terminating at the northern bow shock where it has strengthened the pre-existing water masers.  We note that a radiative origin for variations in water maser intensity has been invoked in the past to explain the observed rapid variations of Cepheus~A and W3(OH) even while the primary pumping mechanism is collisional in nature \citep{Xiang91,Rowland86,Burke78}.

In a recent 1.3 to 6~cm survey of jets from young massive protostars, 10 of 26 ionized jets showed areas of non-thermal emission in their lobes (spectral index $\approx -0.55$), which was interpreted as Fermi acceleration in shocks \citep{Purser16}.  Thus, the existence of areas of shock-induced emission like CM2 is not uncommon \citep[also see][]{RodriguezKamenetzky17}. We also note that the projected separation of CM2 from the central driving source (MM1B) is $\sim 3000$~au, which is strikingly similar to the projected separation of the masers found in the recent Orion~KL outburst from the protostar Source I \citep[3400~au,][]{Hirota11}.  This agreement suggests a similar combination of evolutionary state and surrounding gas density structures for these two massive protostars.

\subsection{Drop in water maser emission toward MM1}
\label{mm1drop}

In general, maser pumping schemes contain radiative and collisional components, though often one dominates. Additionally, it is important to recognize that while water maser transitions are in the centimeter to submillimeter regime, many of the water transitions that lead to the overpopulation of the masing states occur in the mid-infrared \citep[e.g.][]{Neufeld91}. 
In the context of an accretion outburst, immediately after the rise in protostellar luminosity, the dust efficiently absorbs continuum radiation  and thus heats up more quickly than the gas, which absorbs only in relatively narrow lines. As a result, the temperature of the grains surrounding the flaring protostar (\tdust) can rise dramatically \citep{Johnstone13}. At this stage, the dust emission considerably increases the density of the infrared radiation field. But since the gas has not heated up yet, the efficiency of the collisional source of pumping initially stays at the same level. The dust plays an important role in the radiative part of the pumping and 
this role is twofold: dust radiation provides the source of photons for the pumping while dust absorption provides the sink for photon energy in the pumping network \citep{Strelnitskii81,Sobolev12}. 
It has been shown that infrared emission alone cannot provide an efficient source for the pumping of the 22~GHz water masers due to energetic considerations \citep{Strelnitskii84,Deguchi81,Shmeld76}.  Thus, it is often stated that collisional pumping dominates for water masers. However, if the dust is too warm to provide an efficient sink for the infrared transitions that take part in the overpopulation of the masing state, then the maser pumping becomes less efficient, regardless of whether the collisional pump remains favorable \citep{Deguchi81,Strelnitskii77}.
For example, calculations by \citet{Yates97} show that water maser gain is reduced by an order of magnitude at \tdust\/=300~K compared to 100~K.
In the context of water masers originating in $J$-shocks, an additional possible effect resulting from an elevated \tdust\/ is that free hydrogen atoms in the post shock gas that encounter dust grains will evaporate from the grains before being able to form molecules \citep{HM79}, which will also reduce the efficiency of the collisional pump. In summary, the rapid rise of the infrared radiation from the central protostar and the subsequent heating of the surrounding dust should result in a net decrease of water maser brightness, which can readily explain the drop we have seen in the central parts of MM1, close to MM1B and MM1D.

In contrast, the 6.7~GHz methanol maser requires
infrared radiation to be pumped along with lower densities in the range of $10^4-10^8$~\ccc\/ \citep{Cragg05}.  Thus, for a region with density satisfying the requirements for both masers ($\sim10^8$~\ccc) and \tdust\/ around 100~K, an
increase in \tdust\/ would favor the 6.7~GHz line and depress the 22~GHz line, while a decrease in \tdust\/ would have the opposite effect.  This scenario thus readily and consistently explains the onset of strong 6.7~GHz masers surrounding MM1 
coupled with the drop in water maser emission close to MM1 as a result of the accretion outburst. Interestingly, recent monitoring of the 22~GHz and 6.7~GHz transitions toward the intermediate mass protostar G107.298+5.639 shows a remarkable pattern of anti-correlated alternating flares of these maser species 
that coincide in position and velocity within 360~au \citep{Szymczak16}, which suggests periodic changes in the radiation field from the central object as a simple explanation (e.g. in contrast to superradiance operating at different timescales \citep{Rajabi17}).

\subsection{Nature of water maser emission toward MM3-UCHII and evidence for bulk motion} \label{bulk}

As demonstrated in Fig.~\ref{MM3}, the UCHII-W1 and UCHII-W2 maser associations appear to have undergone bulk motion to the south-southeast between 2011.7 and 2017.0.  Between these two epochs, we find a separation of $\Delta$~R.A.$=+0\farcs036\pm0\farcs01$ and $\Delta$~Dec.$=-0\farcs09\pm0\farcs01$, and a position angle of $-22.1\arcdeg$. These maser associations are coincident with the southernmost CS (6-5) knot in the N-S outflow (see Fig.~\ref{allmasers}).  Strong evidence for the persistent motion of a complex association of water masers has been reported in other protostellar outflows \citep[e.g.,][]{Goddi06,Gallimore03}. 
If we interpret the shift in position of the maser associations between epochs as bulk motion over the intervening 5.33~yr, then the implied vector sum proper motion on the sky plane is  $0.0182\pm0.0019~\arcsec$~yr$^{-1}$, in turn 
yielding a sky plane velocity estimate of $112\pm12$~\kms\/ and a dynamical time of $170\pm18$~yr, consistent with a high velocity outflow.  The most blueshifted masers in UCHII-W1 and UCHII-W2 have  velocities of $-39.0$ and $-40.25$~\kms, respectively, yielding a mean maximum  radial velocity of 32.6~\kms\/ with respect to the systemic velocity ($-7$~\kms\/) and MM1B.  The inferred inclination angle of the maser motion, $\theta_{\rm H_2O}$, derived from the two component velocities is then ${\tt arctan}(32.6/112)=16\pm2\arcdeg$,
and the 3D vector velocity is $v_{\rm jet}=117\pm12$~\kms.
In \S\ref{outflow}, we estimated a 
dynamical time for the N-S outflow of $\sim 580\,\tan(\theta_{inc})$, which, by 
assuming $\theta_{inc} = \theta_{H_2O}$, we can now refine to a dynamical time 
estimate of $\sim 166\pm22$~yr.

In light of this result, we also searched for bulk motion between epochs toward the CM2-W1 bow shock.  The masers at the apex have moved northward by no more than $0\farcs02 = 26$~au, which is similar to the relative uncertainty in the fitted positions.  This upper limit yields a proper motion upper limit of $<3.8$~mas~yr$^{-1}$, which corresponds to a sky plane velocity of $<23$~\kms\/ and (correcting for the outflow inclination angle $\theta_{inc}$) a bow shock advancement velocity ($v_{\rm bs}$) of $<24$~\kms.  The ratio of $v_{\rm jet}/v_{\rm bs}$ is then $>5$, which implies a density ratio between the jet and the ambient medium of $\eta < 0.066$ \citep[see Eq. 1 of][]{Blondin90}. Thus, the jet is a ``light'' ($\eta << 1$), high Mach number jet, consistent with other observed protostellar jets \citep{RodriguezKamenetzky17,Gomez99,Devine97} and the simulations of \citet{Downes03}.

Interestingly, although the excited OH 6.035~GHz masers have long been thought to be related to the MM3-UCHII region,  the UCHII-OH5, UCHII-OH6, UCHII-OH7 maser associations are also coincident with the southern end of the N-S outflow as traced by CS(6-5) (see Fig.~\ref{allmasers}a,c). These three maser associations are also precisely the ones that show a reversal of the Zeeman $B_{los}$ magnetic field direction compared to all the masers coincident with the more southerly parts of the UCHII region, as reported by \citet{Hunter18} and \citet{Caswell11}. Moreover, the UCHII-OH7 maser association is in close proximity ($< 100$~au in projection) to the water masers of the UCHII-W2 association. Unfortunately, too few excited OH 6.035~GHz masers are detected in the \citet{Caswell11} data (epoch 2001.0) to discern if the OH masers show the same bulk motion as the water maser association UCHII-W2. Together, these clues suggest that, like the water masers, the UCHII-OH5, UCHII-OH6, and UCHII-OH7 maser associations are excited by shocks within the outflow, or possibly by interaction between the southern lobe of the outflow and the UCHII region. 


\subsection{The nature of the southernmost water maser \ngci-South}

The \ngci-South masers were detected in both the 2017.0 and 2017.8 water maser epochs. We searched SIMBAD for counterparts to \ngci-South (Table~\ref{waterassoc}), and the only nearby object is the X-ray source CXOU 172053.21-354726.4, which is offset $2\farcs7$ north of the maser. The X-ray source has been identified as a possible counterpart to the infrared star TPR-9 \citep{Tapia96} as their respective positions are within $1''$.   It is possible that the maser is associated with an outflow from this star.  We note that the separation from MM1B is $31\farcs8$, which corresponds to a projected distance of 239~light days. Since our VLA maser observations occurred more than this many days after the origin of the outburst \citep{Macleod18}, we cannot exclude the  possibility that this maser was somehow pumped by narrowly beamed radiation from MM1B.

\section{Conclusions}

The ongoing accretion outburst from \ngci-MM1B provides an important clarifying view into the complex observed phenomena of massive star formation.  In this paper, we have presented the detailed changes in water maser emission in 3 epochs of VLA data from before and after the onset of the 2015 outburst.  The masers in the vicinity of MM1B and MM1D have faded substantially, by factors of 5 to 16, while some velocity components have disappeared entirely, which we attribute to the increased dust temperature reducing the efficiency of the maser pump.   In the most recent epoch, a weak redshifted component has appeared northeast of MM1D and coincides with a weak extension of the 6~cm jet driven by MM1B.  Further north along the jet axis at the synchrotron source CM2, spectacularly flaring water masers (mean factor of 6.5) trace a remarkably complete and symmetric bow shock pattern.  The previously-observed excited-state OH masers trace a similar bow shock pattern located 120~au downstream in cooler and slightly less dense gas.  

In ALMA images of the dense thermal gas tracer CS (6-5), we identify several collimated outflows, including a newly-recognized, dynamically young, compact north-south bipolar structure emanating from the outburst source MM1B that encompasses the majority of the water and excited OH masers. We conclude that this previously-excavated cavity allowed the radiative energy of the current outburst to propagate freely until terminating at the northern bow shock where it has strengthened the pre-existing water masers. Changes in the water masers toward the southern lobe of the outflow are less dramatic, but there appears to be bulk motion occurring in two of the water maser associations with a proper motion of $0\farcs0182\pm0\farcs0019$~yr$^{-1}$ pointing directly away from MM1B, leading to a dynamical time of $\sim 166$ yr and a velocity of 117~\kms\/ for the MM1B N-S outflow. 

In CS (6-5) we also detect a single-lobed blueshifted outflow to the NW of MM1B as well as the inner regions of the well-known large-scale NE-SW outflow, bringing the total number of outflows from this massive protostar to at least three, suggesting a dynamic picture of evolution in which the outflow orientation can change by large angles over relatively short periods as reported in other protostars including Cepheus~A~HW2 \citep{Cunningham09}, Orion Source~I \citep{Plambeck09}, and the FU~Ori object V1647~\added{Ori} \citep{Principe18}. An alternative interpretation is that these outflows arise from a binary or higher-order multiple system with non-coplanar disks still unresolved within MM1B, such as the disk pair found in IRAS~17216-3801 \citep{Kraus17}.  We have also detected a highly collimated SE-NW outflow toward the enigmatic line-free, but high dust brightness temperature \citep[$\sim 80$~K,][]{Brogan16} millimeter dust core MM4, confirming that it must harbor a protostar. 

Future monitoring of both the millimeter and maser emission toward this protocluster will continue to reveal clues about the dynamic evolution of massive protostars. Upcoming ALMA matched spatial-scale, multi-transition line observations of the outflows will also allow modeling of the physical conditions in the various outflow structures in detail. Although pre-outburst observations of the various submillimeter water masers \citep{Melnick93,Menten91,Menten90} do not exist for this object, future studies to search for these masers and pinpoint their locations would be helpful to better understand the maser  
pumping schemes. Finally, Very Long Baseline interferometry will provide valuable insight about the proper motions of the various maser species in order to better understand the kinematics and time-evolution of the N-S outflow. 

\acknowledgments

The National Radio Astronomy Observatory is a facility of the National Science Foundation operated under agreement by the Associated Universities, Inc.  This paper makes use of the following ALMA data: ADS/JAO.ALMA\#2015.A.00022.T. ALMA is a partnership of ESO (representing its member states), NSF (USA) and NINS (Japan), together with NRC (Canada) and NSC and ASIAA (Taiwan) and KASI (Republic of Korea), in cooperation with the Republic of Chile. The Joint ALMA Observatory is operated by ESO, AUI/NRAO and NAOJ.   C.J.~Cyganowski acknowledges support from the STFC (grant number ST/M001296/1).  T. Hirota is supported by the MEXT/JSPS KAKENHI
grant No. 17K05398. Support for B.A.M. was provided by NASA through Hubble Fellowship grant \#HST-HF2-51396 awarded by the Space Telescope Science Institute, which is operated by the Association of Universities for Research in Astronomy, Inc., for NASA, under contract NAS5-26555. \added{A. M. Sobolev acknowledges support from the Russian Science Foundation (grant number 18-12-00193).} This research made use of NASA's Astrophysics Data System Bibliographic Services, the SIMBAD database operated at CDS, Strasbourg, France,  Astropy, a community-developed core Python package for Astronomy \citep{astropy}, and APLpy \citep{aplpy2012}, an open-source plotting package for Python hosted at http://aplpy.github.com.

\facilities{ALMA, VLA}.

\added{\software{CASA (McMullin et al. 2007), astropy (The Astropy Collaboration 2013), APLpy (Robitaille and Bressert, 2012)}}

\clearpage


\begin{thebibliography}{}

\bibitem[Abraham et al.(1981)]{Abraham81} Abraham, Z., Cohen, N.~L., Opher, R., Raffaelli, J.~C., \& Zisk, S.~H.\ 1981, \aap, 100, L10 


\bibitem[Astropy Collaboration et al.(2013)]{astropy} 
{Astropy Collaboration, Robitaille, T.~P., Tollerud, E.~J., et al.\ 2013, \aap, 558, A33 }





\bibitem[Beuther et al.(2008)]{Beuther08} Beuther, H., Walsh, A.~J., Thorwirth, S., et al.\ 2008, \aap, 481, 169 


\bibitem[Blondin et al.(1990)]{Blondin90} Blondin, J.~M., Fryxell, B.~A., \& Konigl, A.\ 1990, \apj, 360, 370 




\bibitem[Breen et al.(2010)]{Breen10} Breen, S.~L., Caswell, J.~L., Ellingsen, S.~P., \& Phillips, C.~J.\ 2010, \mnras, 406, 1487 

\bibitem[Brogan et al.(2016)]{Brogan16} Brogan, C.~L., Hunter, T.~R., Cyganowski, C.~J., et al.\ 2016, \apj, 832, 187


\bibitem[Burke et al.(1978)]{Burke78} Burke, B.~F., Giuffrida, T.~S., \& Haschick, A.~D.\ 1978, \apjl, 226, L21 

\bibitem[Burns et al.(2017)]{Burns17} Burns, R.~A., Handa, T., Imai, H., et al.\ 2017, \mnras, 467, 2367 

\bibitem[Burns et al.(2016)]{Burns16} Burns, R.~A., Handa, T., Nagayama, T., Sunada, K., \& Omodaka, T.\ 2016, \mnras, 460, 283 

\bibitem[Caratti o Garatti et al.(2017)]{Caratti17} Carotti o Garatti, A., Stecklum, B., Garcia Lopez, R., et al.\ 2017, Nature Physics, 13, 276


\bibitem[Caswell et al.(2011)]{Caswell11} Caswell, J.~L., Kramer, B.~H., \& Reynolds, J.~E.\ 2011, \mnras, 414, 1914 





\bibitem[Chen et al.(2016)]{Chen16} Chen, X., Arce, H.~G., Zhang, Q., Launhardt, R., \& Henning, T.\ 2016, \apj, 824, 72 

\bibitem[Chibueze et al.(2014)]{Chibueze14}
Chibueze, J.~O., Omodaka, T., Handa, T., et al.\ 2014, \apj, 784, 114 

\bibitem[Chibueze et al.(2012)]{Chibueze12} Chibueze, J.~O., Imai, H., Tafoya, D., et al.\ 2012, \apj, 748, 146 




\bibitem[Cragg et al.(2005)]{Cragg05} Cragg, D.~M., Sobolev, A.~M., \& Godfrey, P.~D.\ 2005, \mnras, 360, 533 

\bibitem[Cragg et al.(2002)]{Cragg02} Cragg, D.~M., Sobolev, A.~M., \& Godfrey, P.~D.\ 2002, \mnras, 331, 521 

\bibitem[Cunningham et al.(2009)]{Cunningham09} Cunningham, N.~J., Moeckel, N., \& Bally, J.\ 2009, \apj, 692, 943 

\bibitem[De Buizer et al.(2002)]{DeBuizer02} De Buizer, J.~M., Radomski, J.~T., Pi{\~n}a, R.~K., \& Telesco, C.~M.\ 2002, \apj, 580, 305 

\bibitem[de Jong(1973)]{deJong73} de Jong, T.\ 1973, \aap, 26, 297 

\added{\bibitem[Deguchi(1981)]{Deguchi81} Deguchi, S.\ 1981, \apj, 249, 145 }

\bibitem[Devine et al.(1997)]{Devine97} Devine, D., Bally, J., Reipurth, B., \& Heathcote, S.\ 1997, \aj, 114, 2095 

\bibitem[Downes \& Cabrit(2003)]{Downes03} Downes, T.~P., \& Cabrit, S.\ 2003, \aap, 403, 135 







\bibitem[Ellingsen et al.(2004)]{Ellingsen04} Ellingsen, S.~P., Cragg, D.~M., Lovell, J.~E.~J., et al.\ 2004, \mnras, 354, 401 

\bibitem[Etoka et al.(2012)]{Etoka12} 
{Etoka, S., Gray, M.~D., \& Fuller, G.~A.\ 2012, \mnras, 423, 647 }


\bibitem[Fischer et al.(2017)]{Fischer17} Fischer, W.~J., Megeath, S.~T., Furlan, E., et al.\ 2017, \apj, 840, 69 




\bibitem[Forster \& Caswell(1999)]{FC99} Forster, J.~R., \& Caswell, J.~L.\ 1999, \aaps, 137, 43 

\bibitem[Frimann et al.(2017)]{Frimann17} Frimann, S., J{\o}rgensen, J.~K., Dunham, M.~M., et al.\ 2017, \aap, 602, A120 


\bibitem[Gallimore et al.(2003)]{Gallimore03} Gallimore, J.~F., Cool, R.~J., Thornley, M.~D., \& McMullin, J.\ 2003, \apj, 586, 306 


\bibitem[Goddi et al.(2006)]{Goddi06} Goddi, C., Moscadelli, L., Torrelles, J.~M., Uscanga, L., \& Cesaroni, R.\ 2006, \aap, 447, L9 

\bibitem[Goedhart et al.(2004)]{Goedhart04} Goedhart, S., Gaylard, M.~J., \& van der Walt, D.~J.\ 2004, \mnras, 355, 553 

\added{\bibitem[Goicoechea et al.(2015)]{Goicoechea2015} Goicoechea, J.~R., Chavarr{\'{\i}}a, L., Cernicharo, J., et al.\ 2015, \apj, 799, 102} 


\bibitem[G{\'o}mez de Castro \& Robles(1999)]{Gomez99} G{\'o}mez de Castro, A.~I., \& Robles, A.\ 1999, \aap, 344, 632 

\bibitem[Gottlieb et al.(2003)]{Gottlieb03} Gottlieb, C.~A., Myers, P.~C., \& Thaddeus, P.\ 2003, \apj, 588, 655 

\bibitem[Gray et al.(2016)]{Gray16} Gray, M.~D., Baudry, A., Richards, A.~M.~S., et al.\ 2016, \mnras, 456, 374 


\bibitem[Gustafsson et al.(2010)]{Gustafsson10} Gustafsson, M., Ravkilde, T., Kristensen, L.~E., et al.\ 2010, \aap, 513, A5 

\bibitem[Gwinn(1994)]{Gwinn94} Gwinn, C.~R.\ 1994, \apj, 429, 241 






\bibitem[Hirota et al.(2014)]{Hirota14} Hirota, T., Tsuboi, M., Kurono, Y., et al.\ 2014, \pasj, 66, 106 

\bibitem[Hirota et al.(2011)]{Hirota11} Hirota, T., Tsuboi, M., Fujisawa, K., et al.\ 2011, \apjl, 739, L59 

\bibitem[Hollenbach et al.(2013)]{Hollenbach13} Hollenbach, D., Elitzur, M., \& McKee, C.~F.\ 2013, \apj, 773, 70 


\bibitem[Hollenbach \& McKee(1979)]{HM79} Hollenbach, D., \& McKee, C.~F.\ 1979, \apjs, 41, 555 


\bibitem[Hsieh et al.(2018)]{Hsieh18} Hsieh, T.-H., Murillo, N.~M., Belloche, A., et al.\ 2018, \apj, 854, 15 

\bibitem[Hunter et al.(2018)]{Hunter18} Hunter, T.~R., et al.\ 2018, \apj, 854, 170

\bibitem[Hunter et al.(2017)]{Hunter17} Hunter, T.~R., Brogan, C.~L., MacLeod, G., et al.\ 2017, \apjl, 837, L29 


\bibitem[Hunter et~al.(2006)]{Hunter06} {Hunter}, T.~R., {Brogan}, C.~L., {Megeath}, S. T., et al., 2006, \apj, 649, 888


\bibitem[Jensen \& Haugb{\o}lle(2018)]{Jensen18} Jensen, S.~S., \& Haugb{\o}lle, T.\ 2018, \mnras, 474, 1176

\bibitem[Johnstone et al.(2013)]{Johnstone13} Johnstone, D., Hendricks, B., Herczeg, G.~J., \& Bruderer, S.\ 2013, \apj, 765, 133 

\bibitem[J{\o}rgensen et al.(2013)]{Jorgensen13} J{\o}rgensen, J.~K., Visser, R., Sakai, N., et al.\ 2013, \apjl, 779, L22 

\bibitem[Kaufman \& Neufeld(1996)]{Kaufman96} Kaufman, M.~J., \& Neufeld, D.~A.\ 1996, \apj, 456, 250 




%

\bibitem[Kraus et al.(2017)]{Kraus17} Kraus, S., Kluska, J., Kreplin, A., et al.\ 2017, \apjl, 835, L5 


 

\bibitem[Larson(2003)]{Larson03} Larson, R.~B.\ 2003, Reports on Progress in Physics, 66, 1651 

\bibitem[Leurini et~al.(2006)]{Leurini06}
Leurini, S., Schilke, P., Parise, B., et al., 2006, \aap, 454, L83


\bibitem[Lique et al.(2006)]{Lique06} Lique, F., Spielfiedel, A., \& Cernicharo, J.\ 2006, \aap, 451, 1125





\bibitem[Lomax \& Whitworth(2018)]{Lomax18} Lomax, O., \& Whitworth, A.~P.\ 2018, \mnras, 475, 1696 

\bibitem[Lucas et al.(2008)]{Lucas08} Lucas, P.~W., Hoare, M.~G., Longmore, A., et al.\ 2008, \mnras, 391, 136 

\bibitem[Macleod et al.(2018)]{Macleod18}
MacLeod, G., Smits, D.P., Goedhart, S., et al., 2018, MNRAS, 478, 1077


\bibitem[McCutcheon et al.(2000)]{McCutcheon00} McCutcheon, W.~H., Sandell, G., Matthews, H.~E., et al.\ 2000, \mnras, 316, 152 

\bibitem[McGuire et al.(2018)]{McGuire18}
McGuire, B. A., Brogan, C. L., Hunter, T. R., et al.\ 2018, \apjl, 863, L35

\bibitem[McGuire et al.(2017)]{McGuire17}
McGuire, B. A., Shingledecker, A. M., Willis, E. R., et al.\ 2017, \apjl, 851, L46

\added{\bibitem[McMullin et al.(2007)]{CASA2007} McMullin, J.~P., Waters, B., Schiebel, D., Young, W., \& Golap, K.\ 2007, Astronomical Data Analysis Software and Systems XVI, 376, 127}

\bibitem[Melnick et al.(1993)]{Melnick93} Melnick, G.~J., Menten, K.~M., Phillips, T.~G., \& Hunter, T.\ 1993, \apjl, 416, L37 

\bibitem[Menten \& Melnick(1991)]{Menten91} Menten, K.~M., \& Melnick, G.~J.\ 1991, \apj, 377, 647 

\bibitem[Menten et al.(1990)]{Menten90} Menten, K.~M., Melnick, G.~J., Phillips, T.~G., \& Neufeld, D.~A.\ 1990, \apjl, 363, L27 

\bibitem[Menten \& Batrla(1989)]{Menten89} Menten, K.~M., \& Batrla, W.\ 1989, \apj, 341, 839 

\bibitem[Meyer et al.(2017)]{Meyer17} Meyer, D.~M.-A., Vorobyov, E.~I., Kuiper, R., \& Kley, W.\ 2017, \mnras, 464, L90 

\bibitem[Minniti et al.(2010)]{Minniti10} Minniti, D., Lucas, P.~W., Emerson, J.~P., et al.\ 2010, New Astronomy, 15, 433 


\bibitem[Moscadelli et al.(2011)]{Moscadelli11} Moscadelli, L., Cesaroni, R., Rioja, M.~J., Dodson, R., \& Reid, M.~J.\ 2011, \aap, 526, A66 

\bibitem[Neufeld \& Dalgarno(1989)]{Neufeld89} Neufeld, D.~A., \& Dalgarno, A.\ 1989, \apj, 340, 869 

\bibitem[Neufeld \& Melnick(1991)]{Neufeld91} Neufeld, D.~A., \& Melnick, G.~J.\ 1991, \apj, 368, 215 

\bibitem[Norman et al.(1982)]{Norman82} Norman, M.~L., Winkler, K.-H.~A., Smarr, L., \& Smith, M.~D.\ 1982, \aap, 113, 285 

\bibitem[Ogbodo et al.(2017)]{Ogbodo17} Ogbodo, C.~S., Burns, R.~A., Handa, T., et al.\ 2017, \mnras, 469, 4788 

\bibitem[Omodaka et al.(1999)]{Omodaka99} Omodaka, T., Maeda, T., Miyoshi, M., et al.\ 1999, \pasj, 51, 333 

\bibitem[Ostriker et al.(2001)]{Ostriker01} Ostriker, E.~C., Lee, C.-F., Stone, J.~M., \& Mundy, L.~G.\ 2001, \apj, 557, 443 

\bibitem[Plambeck et al.(2009)]{Plambeck09} Plambeck, R.~L., Wright, M.~C.~H., Friedel, D.~N., et al.\ 2009, \apjl, 704, L25 

\bibitem[Plunkett et al.(2015)]{Plunkett15} Plunkett, A.~L., Arce, H.~G., Mardones, D., et al.\ 2015, \nat, 527, 70 

\bibitem[Principe et al.(2018)]{Principe18} Principe, D.~A., Cieza, L., Hales, A., et al.\ 2018, \mnras, 473, 879 

\bibitem[Purser et al.(2016)]{Purser16} Purser, S.~J.~D., Lumsden, S.~L., Hoare, M.~G., et al.\ 2016, \mnras, 460, 1039 

\bibitem[Qiu et al.(2011)]{Qiu11} Qiu, K., Wyrowski, F., Menten, K.~M., et al.\ 2011, \apjl, 743, L25 

\bibitem[Qiu \& Zhang(2009)]{Qiu09} Qiu, K., \& Zhang, Q.\ 2009, \apjl, 702, L66 

\bibitem[Rajabi \& Houde(2017)]{Rajabi17} Rajabi, F., \& Houde, M.\ 2017, Science Advances, 3, e1601858 

\bibitem[Reid et al.(2014)]{Reid14} Reid, M.~J., Menten, K.~M., Brunthaler, A., et al.\ 2014, \apj, 783, 130 

\added{\bibitem[Robitaille \& Bressert(2012)]{aplpy2012} Robitaille, T., \& Bressert, E.\ 2012, Astrophysics Source Code Library, ascl:1208.017 }


\bibitem[Rodr{\'{\i}}guez-Kamenetzky et al.(2017)]{RodriguezKamenetzky17} Rodr{\'{\i}}guez-Kamenetzky, A., Carrasco-Gonz{\'a}lez, C., Araudo, A., et al.\ 2017, \apj, 851, 16 

\bibitem[Rodriguez et al.(1982)]{Rodriguez82} Rodriguez, L.~F., Canto, J., \& Moran, J.~M.\ 1982, \apj, 255, 103 

\bibitem[Rowland \& Cohen(1986)]{Rowland86} Rowland, P.~R., \& Cohen, R.~J.\ 1986, \mnras, 220, 233 

\bibitem[Rupen(1999)]{Rupen99} Rupen, M.~P.\ 1999, Synthesis Imaging in Radio Astronomy II, ASP Conf. Series 180, 229 

\bibitem[Safron et al.(2015)]{Safron15} Safron, E.~J., Fischer, W.~J., Megeath, S.~T., et al.\ 2015, \apjl, 800, L5 




\bibitem[Sault \& Sowinski(2013)]{Sault13} Sault, R.J. \& Sowinski, K.\ 2013, EVLA Memo 169

\added{\bibitem[Shmeld(1976)]{Shmeld76} Shmeld, I.~K.\ 1976, \sovast, 20, 571}

\bibitem[Sch{\"o}ier et al.(2005)]{Schoier05} Sch{\"o}ier, F.~L., van der Tak, F.~F.~S., van Dishoeck, E.~F., \& Black, J.~H.\ 2005, \aap, 432, 369



\added{\bibitem[Sobolev \& Gray(2012)]{Sobolev12} Sobolev, A.~M., \& Gray, M.~D.\ 2012, Cosmic Masers - from OH to H0, IAU Symposium 287, 13 }


\bibitem[S{\'a}nchez-Monge et al.(2018)]{Sanchez18} S{\'a}nchez-Monge, {\'A}., Schilke, P., Ginsburg, A., Cesaroni, R., \& Schmiedeke, A.\ 2018, \aap, 609, A101 

\bibitem[Stamatellos et al.(2012)]{Stamatellos12} Stamatellos, D., Whitworth, A.~P., \& Hubber, D.~A.\ 2012, \mnras, 427, 1182 

\bibitem[Stecklum et al.(2017)]{Stecklum17} Stecklum, B., Garatti, A.~C.~o, Hodapp, K., et al.\ 2017, arXiv:1711.01489 


\added{\bibitem[Strelnitskii(1984)]{Strelnitskii84} Strelnitskii, V.~S.\ 1984, \mnras, 207, 339 }

\added{\bibitem[Strelnitskii(1981)]{Strelnitskii81} Strelnitskii, V.~S.\ 1981, \sovast, 25, 373 }

\added{\bibitem[Strelnitskii(1977)]{Strelnitskii77} Strelnitskii, V.~S.\ 1977, \sovast, 21, 381 }

\bibitem[Szymczak et al.(2016)]{Szymczak16} Szymczak, M., Olech, M., Wolak, P., Bartkiewicz, A., \& Gawro{\'n}ski, M.\ 2016, \mnras, 459, L56 


\bibitem[Tapia et al.(1996)]{Tapia96} Tapia, M., Persi, P., \& Roth, M.\ 1996, \aap, 316, 102 



\bibitem[Taylor et al.(1992)]{Taylor92} Taylor, D., Dyson, J.~E., \& Axon, D.~J.\ 1992, \mnras, 255, 351 


\bibitem[Titmarsh et al.(2016)]{Titmarsh16} Titmarsh, A.~M., Ellingsen, S.~P., Breen, S.~L., Caswell, J.~L., \& Voronkov, M.~A.\ 2016, \mnras, 459, 157 

\bibitem[Tofani et al.(1995)]{Tofani95} Tofani, G., Felli, M., Taylor, G.~B., \& Hunter, T.~R.\ 1995, \aaps, 112, 299 


\bibitem[Tolmachev(2011)]{Tolmachev11} Tolmachev, A.\ 2011, The Astronomer's Telegram, 3177

\bibitem[Torrelles et al.(2014)]{Torrelles14} Torrelles, J.~M., Curiel, S., Estalella, R., et al.\ 2014, \mnras, 442, 148 






\bibitem[Vorobyov \& Basu(2005)]{Vorobyov05} Vorobyov, E.~I., \& Basu, S.\ 2005, \apjl, 633, L137 






\bibitem[Xiang et al.(1991)]{Xiang91} Xiang, D., Tang, Y., \& Yu, Z.\ 1991, \apss, 186, 21 

\bibitem[Yates et al.(1997)]{Yates97} Yates, J.~A., Field, D., \& Gray, M.~D.\ 1997, \mnras, 285, 303 

\bibitem[Yoo et al.(2017)]{Yoo17} Yoo, H., Lee, J.-E., Mairs, S., et al.\ 2017, \apj, 849, 69



\bibitem[Zernickel et al.(2012)]{Zernickel12} Zernickel, A., Schilke, P., Schmiedeke, A., et al.\ 2012, \aap, 546, A87 




\end{thebibliography}
\end{document}